\documentclass[10pt,twocolumn,twoside]{IEEEtran}   
\pdfoutput=1

\usepackage{t1enc}
\usepackage{amsmath}    
\usepackage{amsfonts}   
\usepackage{amssymb}
\usepackage{graphicx}
\usepackage{epstopdf}
\usepackage{color}

\usepackage{cite}

\usepackage{multirow}

\usepackage{fancyhdr}
\newcommand{\vol}{$\mathcal{V}$ }

\usepackage{array}
\newcolumntype{L}[1]{>{\raggedright\let\newline\\\arraybackslash\hspace{0pt}}m{#1}}
\newcolumntype{C}[1]{>{\centering\let\newline\\\arraybackslash\hspace{0pt}}m{#1}}
\newcolumntype{R}[1]{>{\raggedleft\let\newline\\\arraybackslash\hspace{0pt}}m{#1}}

\usepackage[tight,footnotesize]{subfigure}
\begin{document}
%
\title{Efficient Steered-Response Power Methods\\ for Sound Source Localization\\ Using Microphone Arrays}
%
%
%

\author{Markus~V.~S.~Lima,
        Wallace~A.~Martins,
        Leonardo~O.~Nunes,
        Luiz~W.~P.~Biscainho,
        Tadeu~N.~Ferreira,
        Maurício~V.~M.~Costa,
        Bowon~Lee
\thanks{Copyright \copyright 2014 IEEE. Personal use of this material is permitted. However, permission to use this material for any other purposes must be obtained from the IEEE by sending a request to pubs-permissions@ieee.org.}
\thanks{M.~V.~S.~Lima, W.~A.~Martins, L.~O.~Nunes, L.~W.~P.~Biscainho, and M.~V.~M.~Costa
are with the Signals, Multimedia and Telecommunications Laboratory, SMT--DEL/Poli \& PEE/COPPE, {Universidade Federal do Rio de Janeiro}, Rio de Janeiro, Brazil
(e-mail: markus.lima@ieee.org, \{wallace.martins, leonardo.nunes, wagner, mauricio.costa\}@smt.ufrj.br).}
\thanks{T.~N.~Ferreira is with the Telecommunications Engineering Department, Fluminense Federal University, Niterói, Brazil
(e-mail: tadeu\_ferreira@id.uff.br).}
\thanks{B.~Lee is with the Department of Electronic Engineering, Inha University, Incheon, South Korea  (e-mail:bowon.lee@inha.ac.kr). He performed the work while at Hewlett-Packard Laboratories.}
}
\maketitle

\begin{abstract} 
This paper proposes an efficient method based on the steered-response power (SRP)
technique for sound source localization using microphone arrays: the volumetric SRP
(V-SRP). As compared to the SRP, by deploying a sparser volumetric grid, the V-SRP
achieves a significant reduction of the computational complexity without sacrificing
the accuracy of the location estimates. By appending a fine search step to the V-SRP,
its refined version (RV-SRP) improves on the compromise between complexity and accuracy.
Experiments conducted in both {simulated}- and real-data scenarios demonstrate the benefits
of the proposed approaches. Specifically, the RV-SRP is shown to outperform the SRP in
accuracy at a computational cost of about ten times lower.
\end{abstract}


%
\IEEEpeerreviewmaketitle

\section{Introduction}\label{sec:intro}

\IEEEPARstart{S}{ound} source localization (SSL) {with microphone arrays} is key to many applications
such as 3-D audio {capture}, speech enhancement for hearing aids in medical applications, vehicle and gunshot
localization for military use, automatic camera steering for event broadcasting or {video conferencing},
and video games~\cite{Benesty_MAbook}.
SSL methods exploit spatial diversity by using {multiple microphones} to simultaneously acquire {different} versions
of emitted source signals, which are then jointly processed.
Knowing the location of a given source enables the enhancement of its associated acquired signals, e.g.
beamforming~\cite{Benesty_MAbook}, thus providing higher signal-to-noise ratio (SNR) than a single-microphone
capture would achieve.

The SSL field has borrowed/extended many of the techniques proposed for source localization
using antenna arrays, which has been an active research area for more than forty years~\cite{Dudgeon_Arraybook}.
In the antenna array framework, most classical algorithms~\cite{UNITARY_ESPRIT,MATRIX_PENCIL,ROOT_MUSIC} were
developed under the assumption that transmitted signals are sufficiently narrowband to allow that phase drifts
between the impinging signals on the several receiving sensors can be attributed only to source positioning.
Since this narrow-band hypothesis does not hold for speech signals, {broadband algorithms} are
the best choice in the SSL scenario~\cite{Handbook_cap51}.

A common approach to solve the localization problem is first estimating the {\it time-differences-of-arrival} (TDoAs) between the acquired signals
and then mapping them into a source position.
When the far-field hypothesis is valid, i.e. the distance between source and array is greater than approximately ten times~\cite{Benesty_MAbook} the length of the array aperture, {the algorithms for DoA (direction-of-arrival) estimation} can be employed.

An intuitive way to estimate the TDoA related to a pair of microphones can be devised if the cross-correlation between their two acquired signals
is known: the lag associated with the maximum measured correlation provides the TDoA estimate itself. This is the basis of the cross-correlation (CC) method
for source localization~\cite{Benesty_MAbook}. The generalized cross-correlation (GCC) method~\cite{CARTER_76} adds robustness to the CC method by including a weighting
function in the cross-spectrum. Different choices for this function lead to different algorithms~\cite{CARTER_73,Brandstein1997,Benesty_MAbook}, among which the {\it phase-transform}
(PHAT) GCC~\cite{CARTER_76} is the preferred scheme.

A natural extension of the GCC technique is the {\it steered-response power} (SRP)~\cite{Omologo1994,Omologo1996,OmologoSvaizerMori1997,DiBiase2000,DiBiase_cap8}
method, which from now on will be denominated {\it classical SRP} (C-SRP).
Compared to the GCC, which first estimates the TDoAs between acquired signals, the C-SRP algorithm becomes more robust to reverberation and noise effects~\cite{Benesty_MAbook,DiBiase_cap8,Silverman2007} by performing a global optimization using all available information. In general terms, the C-SRP method
can be implemented in two steps: (i) compute the cross-correlation function between the signals acquired by each microphone pair; and (ii) search for the source location
over a grid of spatial points.
The second stage is usually the most computationally demanding one since high localization accuracy
implies using dense grids, which can be a major problem especially when facing large search spaces.

In addition to the computational problem due to the use of dense grids, {increasing
the number of microphones within an array has also been used to increase accuracy when the target application allows,
as in the huge microphone arrays presented in~\cite{Silverman1998_HugeMA,Tamai2004_HugeMA}.}
The number of grid points as well as the number of microphones impact directly the computational burden of the C-SRP,
which even under common practical situations may not achieve the desired accuracy. Therefore, SSL solutions
aiming at reducing the computational complexity of SRP-based methods are called~for.

%


\subsection{Main Contributions}

This paper proposes an efficient SRP-based procedure to tackle the problem of sound source localization.
In contrast to the point-wise C-SRP method, the proposed {\it volumetric SRP} (V-SRP) algorithm operates on
pre-defined non-overlapping spatial regions (hereafter loosely called {\it volumes}), each one containing a set
of two or more grid points. The fact that there are more grid points than volumes, which then define
a {\it volumetric grid}, is explored to significantly reduce the computational cost of the method,
as it will become clear in Section~\ref{sec:num_op}. The V-SRP method has proven to be effective
even when using sparse volumetric grids, i.e. large volumes.

Additionally, a {\it refined V-SRP} (RV-SRP) method is devised to tackle those situations when high accuracy is
extremely important. In fact, it looks for a compromise between computational complexity and localization accuracy.
This alternative method departs from the V-SRP search result, over which it performs
a second refining step.
The overall search results are much more accurate at little additional cost over the V-SRP.

Besides, both V-SRP and RV-SRP focus on reducing the number of computations required by the second step inherent to
most SRP-based techniques, namely the search stage. {The idea of a volumetric SRP was originally proposed in~\cite{Said2009} using a different formulation.}

\subsection{Related Works}

Several other methods have also been proposed to reduce the computational complexity of the C-SRP.
In~\cite{Silverman2007}, for example, an improved search method for the C-SRP was proposed, where
Eq.~\eqref{eq:srp_cost_func} is employed, but the stochastic region contraction (SRC) algorithm is
used to find the source position without having to evaluate the {objective function} for every grid point.
This method was then further improved by the use of particle filters in~\cite{Do2009}.
In~\cite{Benesty2007}, a two-step approach is employed in order to reduce the computational complexity of
the C-SRP method.
In particular, only the TDoAs  associated with high-energy cross-correlation values are considered in the search,
thus reducing the computational complexity.

The previous methods tried to reduce the computational complexity by avoiding the computation of the
{objective function} for every point in the search grid.
The approach employed by the V-SRP proposed in this paper is to obtain an SRP-based method for volumetric regions, allowing for the use
of sparser search grids without {compromising its performance}.
If a more precise estimate of the source position is needed, then a second (low-cost) stage can be employed (RV-SRP).

As mentioned before, the original proposal of an SRP operating on volumetric regions is described in~\cite{Said2009}.
In addition to their searching process strategies, another key difference between the proposed V-SRP/RV-SRP and~\cite{Said2009}
lies in their objective functions.
While the objective function of~\cite{Said2009} performs an accumulation of the energy of each point inside a volume,
the proposed algorithm performs this accumulation over the TDoAs associated with the volume.

The algorithm proposed in~\cite{Cobos_letter2011} uses an objective function similar to the one proposed in this paper.
In Section~\ref{sub:diffs}, the differences between the two approaches are detailed.


\subsection{Organization}

This paper is organized as follows.
Section~\ref{sec:csrp} reviews the C-SRP algorithm.
In Section~\ref{sec:vsrp} the V-SRP and RV-SRP methods are proposed.
{Section~\ref{sec:cobos} discusses the algorithm proposed in~\cite{Cobos_letter2011}, focusing on similarities
and differences with respect to the proposed approach.}
Implementation aspects of the aforementioned techniques are addressed in Section~\ref{sec:implementation}.
In Section~\ref{sec:num_op}, assuming a given cost-reducing strategy is employed, the number of arithmetic operations required by each method is computed.
Simulation results for {simulated}- and real-data scenarios demonstrating the good performance of the proposed methods are shown in Section~\ref{sec:results}.
Conclusions are drawn in Section~\ref{sec:conclusion}.

\subsection{Notation}

The symbols $\mathbb{R}$, $\mathbb{Z}$, and $\mathbb{N}$ denote the field of real numbers, the set of integer numbers,
and the set of natural numbers, respectively.
The set of non-negative real numbers is represented by $\mathbb{R}_+$.
In addition, vectors are denoted by lowercase boldface letters, $\| \cdot \|$ is the Euclidean norm,
and the symbol $\lfloor (\cdot) \rfloor$ represents
the highest integer number that is smaller than or equal to the argument $(\cdot)$ (floor operator).

\section{Classical SRP Method} \label{sec:csrp}
The main idea behind the C-SRP method is to steer the array directionality
pattern to different regions, searching for the acoustic source position which is indicated by the maximum power
of the array output signal.
Mathematically, the goal of the C-SRP is to find a point ${\bf x} = (x,y,z) \in \mathbb{R}^3$
that maximizes the {objective function} $W ({\bf x}) \in \mathbb{R}$ given by
\begin{align} \label{eq:srp_cost_func}
 W({\bf x}) \triangleq \sum_{p=1}^P\phi_{p}[{\zeta}_{p}({\bf x})],
\end{align}
in which $P \in \mathbb{N}$ denotes the number of distinct microphone pairs in the array, i.e.
\begin{align}\label{eq:numMicPairs}
 P \triangleq \frac{M (M-1)}{2},
\end{align}
where $M \in \mathbb{N}$ represents the number of microphones.
In addition, ${\zeta}_{p} ({\bf x}) \in \mathbb{Z}$ is defined as
\begin{align} \label{eq:xi_p}
 {\zeta}_{p}({\bf x}) \triangleq
               {\rm round}\left\{ \frac{\| {\bf m}_{p,2} - {\bf x} \| - \| {\bf m}_{p,1} - {\bf x} \|}{c} f_s \right\},
\end{align}
that is, ${\zeta}_{p}({\bf x})$ represents the TDoA {(measured in samples)} from point ${\bf x}$ to the microphone locations ${\bf m}_{p,1}, {\bf m}_{p,2} \in \mathbb{R}^3$,
in which $f_s, c \in \mathbb{R}_+$ denote the sampling rate and the propagation {speed} of sound, respectively.
Denoting the signals acquired by the first and second microphones of the $p$th microphone pair
by $s_{p,1}[n]$ and $s_{p,2}[n]$, the function $\phi_{p} [\zeta] \in \mathbb{R}$ is defined as
\begin{align} \label{eq:phi_p}
 \phi_{p}[\zeta] \triangleq \sum_{n \in \mathbb{Z}} s_{p,1}[n] s_{p,2}[n-\zeta] .
\end{align}
Hence, $\phi_{p}[{\zeta}_{p}({\bf x})]$ is the measured cross-correlation function between the signals acquired by the
$p$th microphone pair for a given TDoA ${\zeta}_{p}({\bf x})$.

Alternatively, $s_{p,1}[n]$ and $s_{p,2}[n]$ may also be filtered versions of the signals recorded by the two microphones
of the $p$th pair. This is the case, for example, when PHAT filtering is used~\cite{Benesty_MAbook,DiBiase2000,DiBiase_cap8}.

\section{Volumetric SRP Method} \label{sec:vsrp}

As mentioned before, the capability of the C-SRP to localize sound sources relies on the assumption that
the acoustic activity at the actual source position is larger than at other positions.
Such acoustic activity is estimated by means of the {objective function} $W({\bf x})$ in Eq.~\eqref{eq:srp_cost_func}, which
is computed based on the TDoAs between the point $\bf x$ and each microphone pair.
{Thus, one can see $W({\bf x})$ as a sum of $ \phi_{p}[\zeta]$ across all microphone pairs, where
for each index $p$ the argument $\zeta$ assumes the value of the $p$th TDoA {as if} the source were at position $\bf x$.
Such objective function, therefore, can be regarded as a
{``soft''} way{\footnote{{Here, the word ``soft'' is employed due to the continuous nature of function $\phi_p$, which is
affected by noise and reverberation effects.}}} of ``counting'' the number of hyperboloids\footnote{{The
term hyperboloid is used in this paper to denote the geometric surface comprised of the points whose associated
TDoAs are equal, i.e., the set $\{ {\bf x} \in \mathbb{R}^3 : {\zeta}_{p}({\bf x}) = \zeta  \}$, where
$\zeta \in \mathbb{Z}$ is a constant.}}
that pass through $\bf x$ and that {are coherent with the source and microphones' positions.}
This {TDoA-counting} process elects the spatial point that maximizes $W({\bf x})$ as the best choice for the source position estimate.}

The idea of the proposed {\it volumetric SRP} (V-SRP) method is to consider spatial information from several points in order
to obtain an estimate of the acoustic activity inside a spatial region.
{Thus, based on the aforementioned reasoning, the goal here is to elect the spatial region corresponding to the maximum value of a similar
{TDoA-counting} process as the one most likely to contain the acoustic source. In this case,
{one should consider all grid points within a volume \vol in order to compute the number of hyperboloids that pass through it.}
By taking into account the spatial information of these points,} the V-SRP will be able to employ sparser volumetric grids,
thus lowering the number of computations without {impacting the performance}. Following the same reasoning of the C-SRP,
the spatial information of the points inside \vol is contained in the TDoAs that they yield.

Mathematically, the V-SRP searches for the spatial region \vol that maximizes the {objective function} $\overline{W}({\cal V})$ given by
\begin{align}\label{eq:vsrp_cost_func}
 \overline{W}({\cal V})
\triangleq \sum_{p=1}^P\sum_{\zeta=\zeta_{p,{\cal V}}^{\rm min}}^{\zeta_{p,{\cal V}}^{\rm
max}}\chi_p[\zeta,{\cal V}] \phi_p[\zeta],
\end{align}
where
\begin{align}\label{eq:vsrp_chi}
 \chi_p[\zeta,{\cal V}] \triangleq \left\{
 \begin{array}{ll}
  1,   &  \text{ if } \zeta = {\zeta}_p({\bf x}) \text{ for some } {\bf x} \in {\cal V} ,  \\
  0,   &  \text{ otherwise. }
 \end{array} \right.
\end{align}
Thus, for a given microphone pair with index $p$ and for a pre-defined volume ${\cal V}$, $\chi_p[\zeta,{\cal V}] = 1$
implies that there exists at least one {grid} point ${\bf x} \in {\cal V}$ such that its associated TDoA ${\zeta}_p({\bf x})$ is
equal to $\zeta$.
In addition, $\phi_p[\zeta]$ is the measured cross-correlation function given by Eq.~\eqref{eq:phi_p},
and $\zeta_{p,{\cal V}}^{\rm min}$ and $\zeta_{p,{\cal V}}^{\rm max} \in \mathbb{Z}$ are the minimum and maximum TDoA
values considering both a specific volume ${\cal V}$ and the $p$th microphone pair.
Hence, one may regard $\chi_p[\zeta,{\cal V}]$ as a selector of lags $\zeta \in \{ \zeta_{p,{\cal V}}^{\rm min},
\ldots , \zeta_{p,{\cal V}}^{\rm max} \} \subset \mathbb{Z}$.
This selector indicates which lags correspond to TDoAs for {grid} points inside the region \vol being evaluated.
After finding the volume ${\cal V}$ that maximizes $\overline{W}({\cal V})$, if one desires a point estimate
for the source location, then the center of the volume may be chosen.\footnote{This is an arbitrary choice that does not have to be made in this particular way.}

\begin{figure}[h!]
\begin{center}
\includegraphics[width=1.2\linewidth]{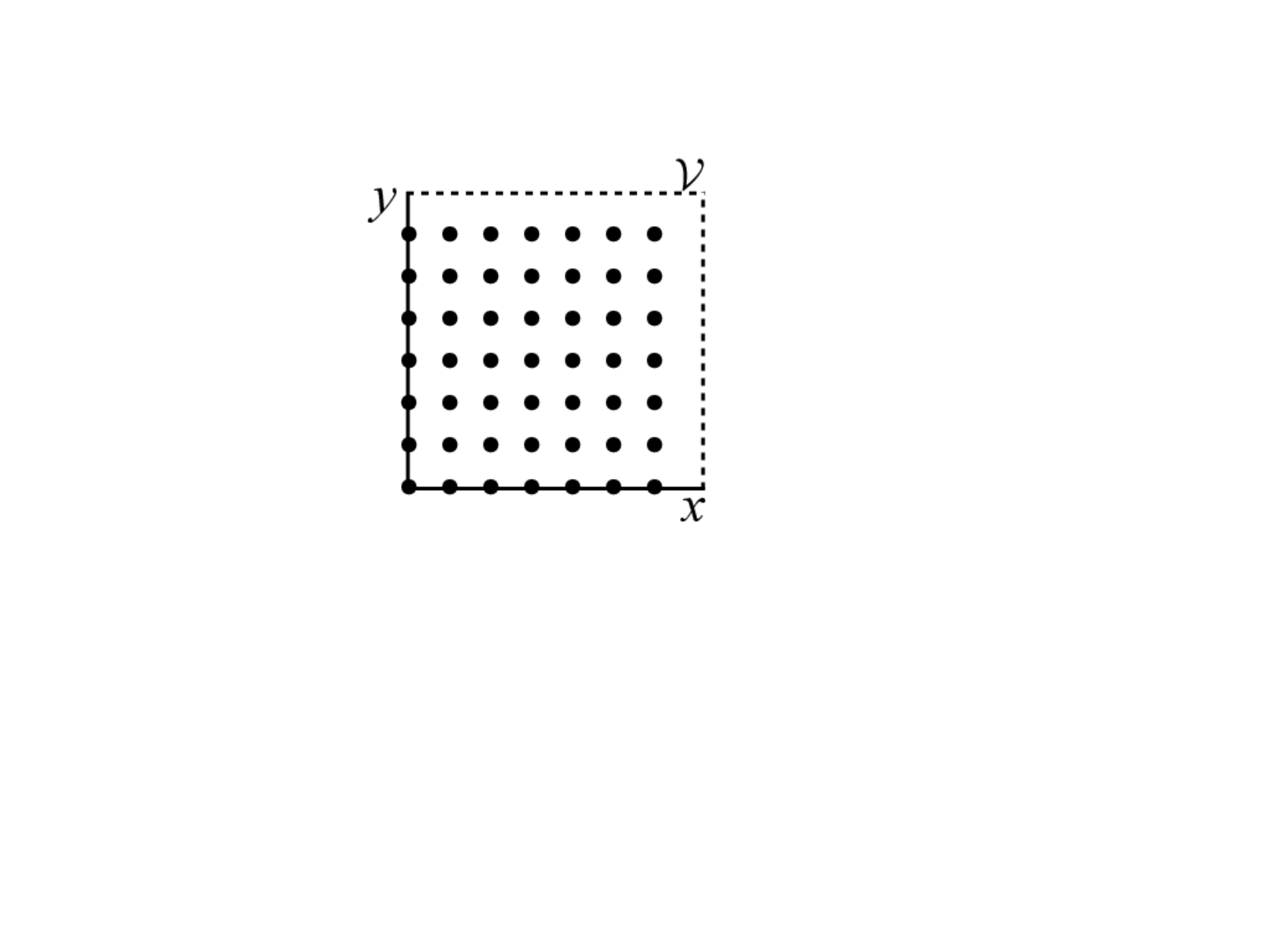} \vspace*{-3.5cm}
\caption{Spatial region \vol with associated points.} \vspace*{-.5cm}
\label{fig:grid}
\end{center}
\end{figure}

{The region \vol} is usually defined as a volume (e.g. a parallelepiped) in a search space contained in $\mathbb{R}^3$,
but if the search space is contained in $\mathbb{R}^2$, then the region degenerates to a plane region (e.g. a rectangle).
In order to allow a simple visualization, Fig.~\ref{fig:grid} shows a set of points inside a given spatial region \vol
contained in $\mathbb{R}^2$, arbitrarily chosen as a square. It is important to highlight that only two edges of the
square are closed, thus indicating that the points on them belong to ${\cal V}$, whereas their opposite edges (in dashed lines)
are open.\footnote{In this example, there are $7^2$ points included in \vol. If it were a cube rather than a
square, \vol would include $7^3$ points, since there would be three concurrent faces closed with their opposite
ones open.} This construction guarantees that there is no {overlap} among adjacent volumes within a
volumetric grid.

\subsection{TDoA Smoothing and Its Spatial Effects} \label{sub:tdoa_smoth}

\begin{figure*}[t!]
\centering
\subfigure[C-SRP: source at position A.]{\includegraphics[scale=0.4]{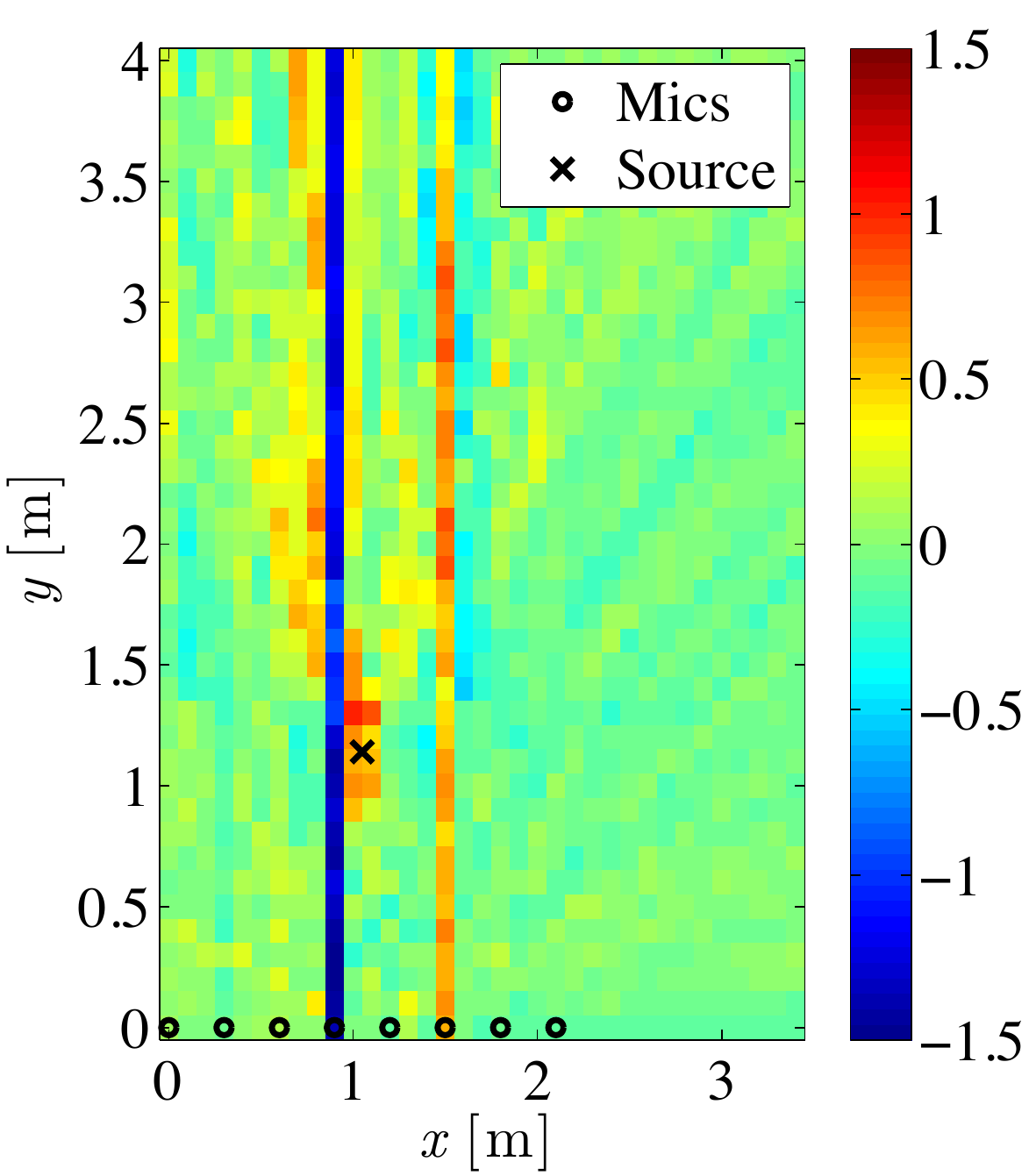}
\label{fig:srp-posA}}
\subfigure[C-SRP: source at position B.]{\includegraphics[scale=0.4]{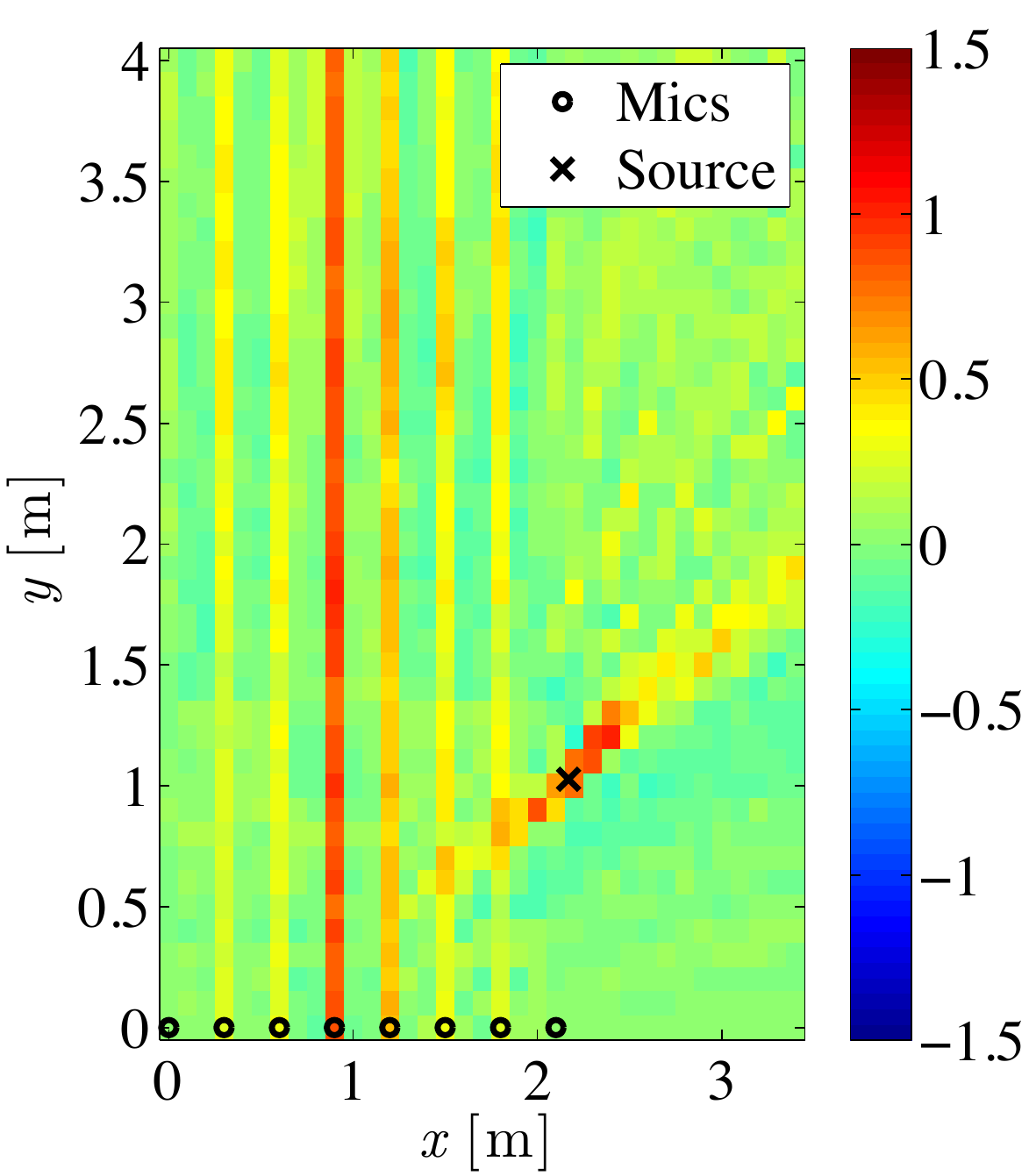}
\label{fig:srp-posB}}
\subfigure[C-SRP: source at position C.]{\includegraphics[scale=0.4]{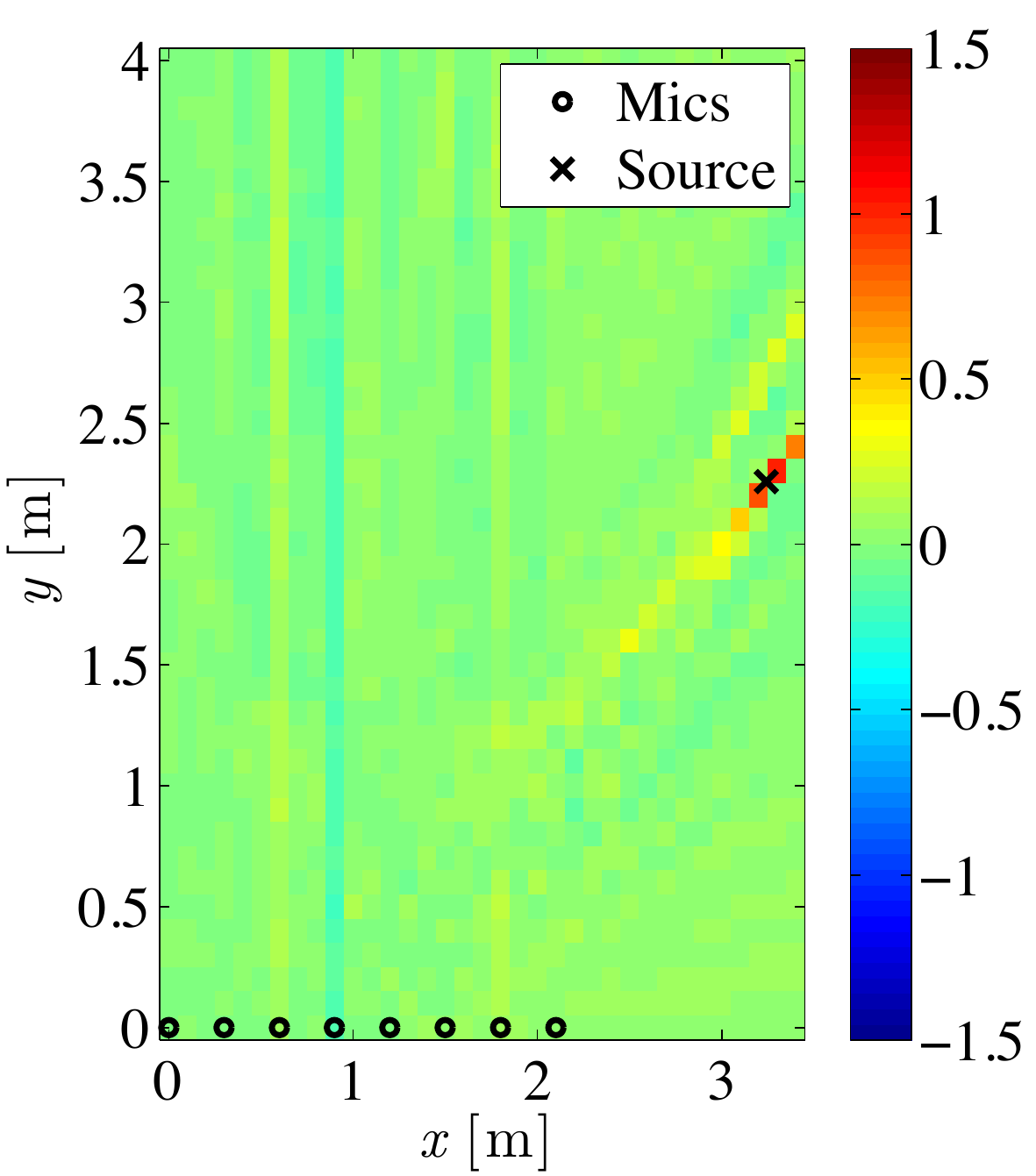}
\label{fig:srp-posC}}\\
\subfigure[V-SRP: source at position A.]{\includegraphics[scale=0.4]{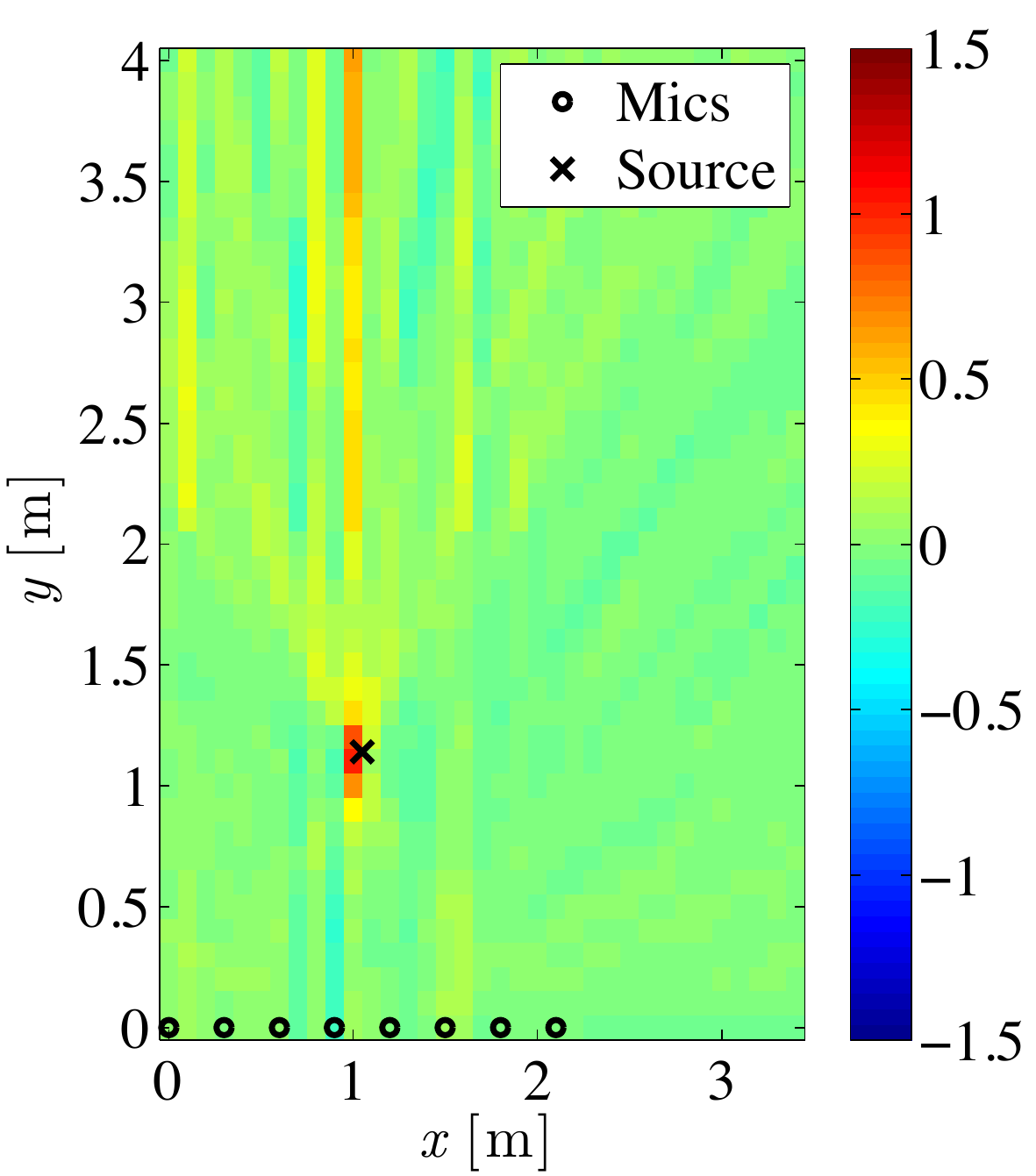}
\label{fig:vol-posA}}
\subfigure[V-SRP: source at position B.]{\includegraphics[scale=0.4]{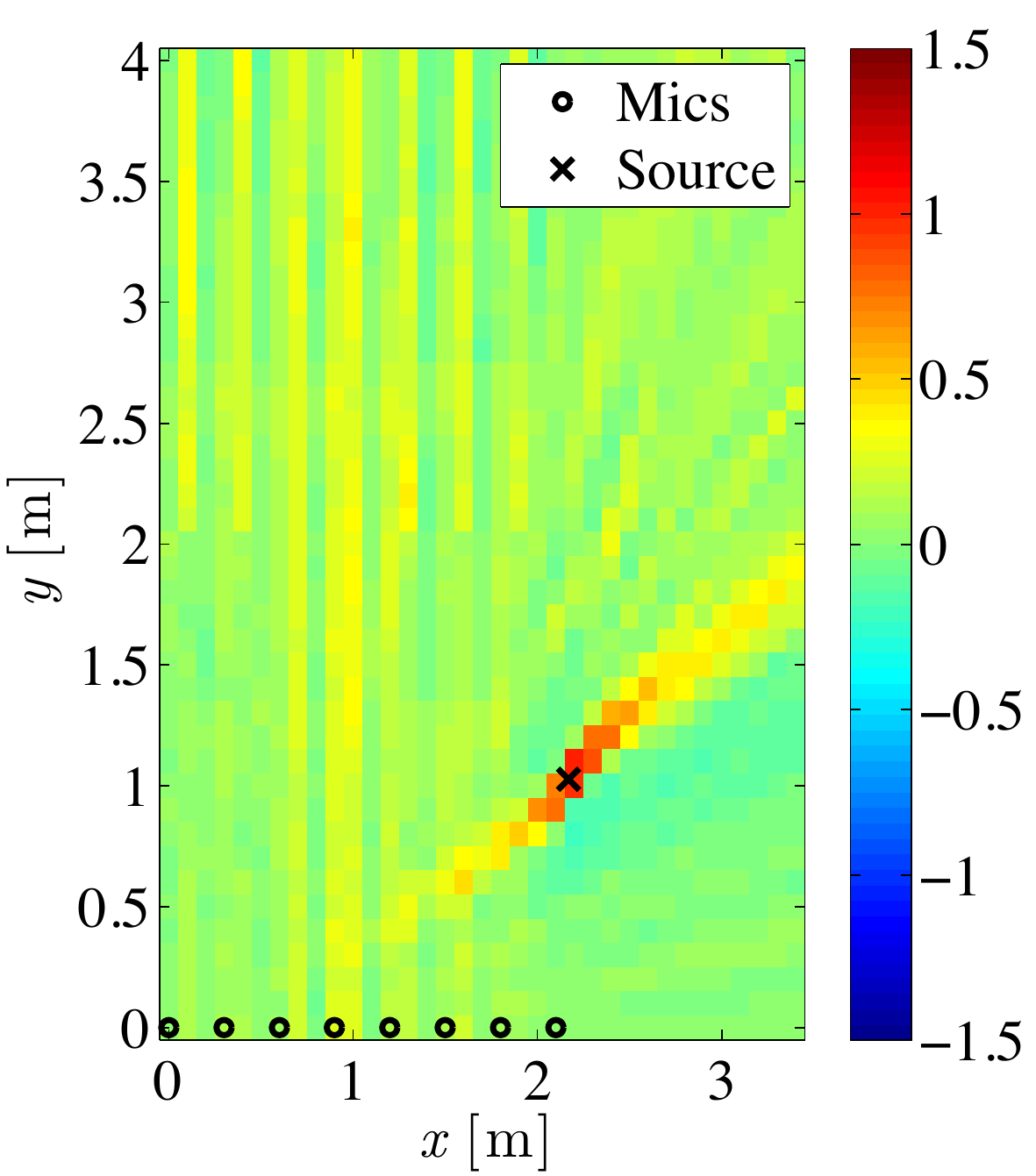}
\label{fig:vol-posB}}
\subfigure[V-SRP: source at position C.]{\includegraphics[scale=0.4]{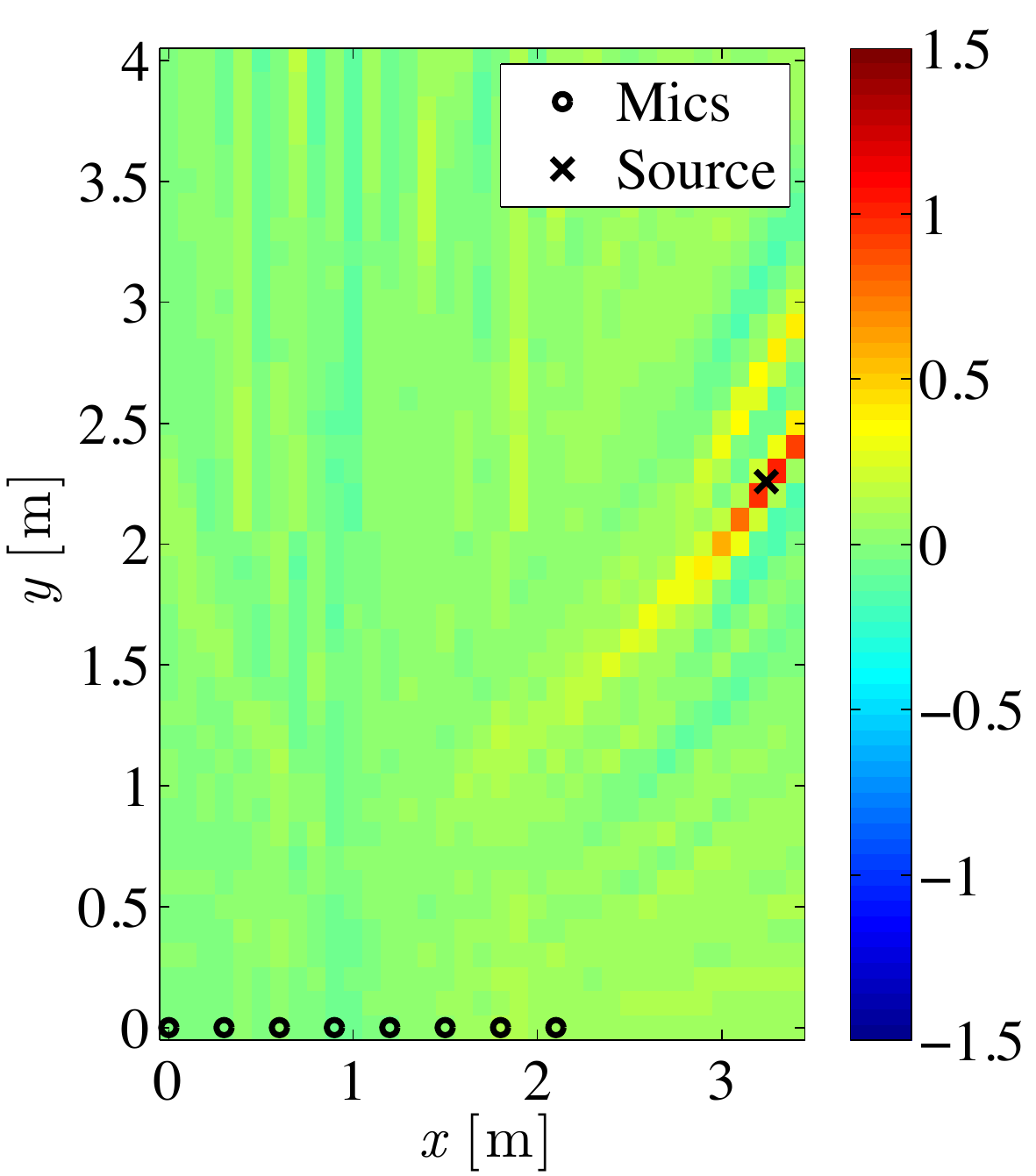}
\label{fig:vol-posC}}
\caption{{Energy maps for C-SRP and V-SRP. Further details about this setup can be found in Subsection~\ref{sub:exper}.}\label{fig:tdoa_smoth}}
\end{figure*}

The V-SRP method in practice employs two grids: one for the spatial regions and another for the points inside each region.
The volumetric grid is employed in the search itself and bounds the accuracy of the method,
whereas the point grid can improve performance by adding more spatial information (through added TDoAs) about a given spatial region.
On the other hand, by increasing the number of points inside a given spatial region, the computational complexity for computing the {acoustic activity}
of a single spatial region is also increased
{(since more terms in $\chi_p$ in Eq.~\eqref{eq:vsrp_chi} are likely to be non-zero)}.
It should be mentioned, however, that since more than one point can be associated with the same delay,
the increase in complexity is not linear with the number of points. Moreover, the indicator function $\chi_p$ can be pre-computed for a given search grid (with associated points) and array geometry, allowing an efficient implementation of the algorithm by avoiding {the multiplications, as will be explained in Section~\ref{sec:implementation}.}

In addition to the sum across all microphone pairs, the presence of another sum operator over the lags in Eq.~\eqref{eq:vsrp_cost_func} indicates that the objective function $\overline{W}({\cal V})$ performs a filtering process along the TDoAs, whose consequent smoothing can mitigate some interferences inherent to the practical TDoA ``counting''. Indeed, as $\phi_p[\zeta]$ in Eq.~\eqref{eq:phi_p} may be affected by reverberation and other acoustic interferences, if one takes into account many lags $\zeta$ instead of just one, then it is likely to achieve a more reliable estimate.

Fig.~\ref{fig:tdoa_smoth} illustrates the spatial effects of the TDoA smoothing considering recorded signals
in a realistic setup described in detail in Subsection~\ref{sub:exper}.
This figure {depicts energy maps} related to C-SRP and V-SRP, which are a pictorial way of representing
$W({\bf x})$ and $\overline{W}({\cal V})$, respectively.
Noting that acoustic sources are indicated by an `x' in the figures, by comparing Fig.~\ref{fig:srp-posA} with
Fig.~\ref{fig:vol-posA} (source at position A) and Fig.~\ref{fig:srp-posB} with Fig.~\ref{fig:vol-posB} (source at position B),
{one can see the advantages of the TDoA smoothing as regards its capability of mitigating
undesirable peaks that are not related to the actual source position.}
Figs.~\ref{fig:srp-posC}~and~\ref{fig:vol-posC} (source at position C) indicate that, when there is no significant secondary peak, the TDoA smoothing does not work against the proposed method.


\subsection{Refined Volumetric SRP Method} \label{sub:eff}

As previously mentioned, the C-SRP method requires dense grids in order to provide accurate estimates of the
source position, especially in reverberant environments.
However, the use of dense grids may be prohibitive in real-time applications and/or for microphone arrays
comprised of a large amount of microphones.

In this context, the V-SRP emerges as a low-cost alternative that provides accurate estimates even when
using sparse volumetric grids in reverberant environments.
Nevertheless, if more accuracy is desired, a refinement stage can be implemented, leading to the
{\it refined volumetric SRP} (RV-SRP) method, which is comprised of two steps:
\begin{enumerate}
 \item First the entire search space is reduced to a volume ${\cal \hat{V}}$, chosen by the V-SRP method;
 \item Considering that the new search space is the volume ${\cal \hat{V}}$, then the C-SRP method is applied inside ${\cal \hat{V}}$ with a dense grid.
\end{enumerate}

The RV-SRP allows one to take advantage of the precise estimation provided by the V-SRP method even when coarse grids are employed,
while obtaining precise point-estimates of the source by using a low-cost  C-SRP due to the limited search region.
The trade-offs as related to computational cost between these methods are detailed in Section~\ref{sec:comp_cost}.

\section{Modified SRP}\label{sec:cobos}

This section describes and compares against the V-SRP a recently proposed SRP-based method whose objective function
looks similar to the one in Eq.~\eqref{eq:vsrp_cost_func}: the modified SRP (M-SRP) method proposed
in~\cite{Cobos_letter2011}.


For a given grid point ${\bf x}$, the {objective function} associated with the M-SRP method depends not
only on the TDoAs from ${\bf x}$ to each pair $p$ of microphones, but also on all other TDoAs related to a cubic volume surrounding ${\bf x}$.
Mathematically, the M-SRP {objective function} can be written as
\begin{equation}\label{eq:cobos}
{W}_{\text{M}}(\mathbf{x}) \triangleq \sum_{p=1}^P\sum_{\zeta = \hat\zeta_{p,{\bf x}}^{{\rm M}, {\rm min}}}^{\hat\zeta_{p,{\bf x}}^{{\rm M}, {\rm max}}}\phi_p[\zeta],
\end{equation}
where the limits of the summation are $\hat\zeta_{p,{\bf x}}^{{\rm M}, {\rm min}} \triangleq \text{round}\left\{ L_{p,1}({\bf x}) f_s \right\}$
and $\hat\zeta_{p,{\bf x}}^{{\rm M}, {\rm max}} \triangleq \text{round}\left\{ L_{p,2}({\bf x}) f_s \right\}$ with the following definitions:
\begin{align}
 L_{p,1}({\bf x}) &\triangleq   \tau_p({\bf x}) - \| \nabla \tau_p({\bf x}) \| d ,                         \\
 L_{p,2}({\bf x}) &\triangleq   \tau_p({\bf x}) + \| \nabla \tau_p({\bf x}) \| d ,                         \\
 \tau_p({\bf x})  &\triangleq   \frac{\| {\bf m}_{p,2} - {\bf x} \| - \| {\bf m}_{p,1} - {\bf x} \|}{c} ,  \\
 d                &\triangleq   \frac{r}{2} \min \left( \frac{1}{|\sin\theta \cos\phi|}, \frac{1}{|\sin\theta \sin\phi|}, \frac{1}{|\cos\theta|}\right) ,
\end{align}
where $r$ is the length of the cube's edge, and $\theta$ and $\phi$ are respectively the elevation and azimuth angles of the gradient $\nabla \tau_p({\bf x})$
in cylindrical coordinates; see Eqs.~(13), (14), and (9) in~\cite{Cobos_letter2011} for more {details.}\footnote{ {While
this paper was under review, a new version of the M-SRP was published in~\cite{Marti2013}. Such new version employs
an iterative search}.}

\subsection{V-SRP vs. M-SRP}\label{sub:diffs}

\begin{table}[!t]
\caption{Summary of differences between V-SRP and M-SRP.}\label{tab:diffs}
\centering
\begin{tabular}{|c|c|c|}
\hline
Differences        &      M-SRP       &   V-SRP   \\
\hline
Grid               &  point grid      &  point and volumetric grids  \\
Volume Shape       &  cubes           &  arbitrary  \\
Lag Weights        &  1 (always)      &  1 or 0  \\
TDoA Bounds        &  may span / extrapolate \vol  &  span \vol \\
\hline
\end{tabular}
\end{table}

In the following, a close comparison between the V-SRP and the M-SRP highlights their differences.
Table~\ref{tab:diffs} summarizes the topics discussed in this subsection.

Although Eqs.~\eqref{eq:vsrp_cost_func} and~\eqref{eq:cobos} are closely related,
they differ from each other in fundamental aspects. They are:
\begin{enumerate}
 \item Grid:
 The M-SRP uses a point grid, as does the C-SRP.
 On the other hand, V-SRP employs two grids, viz. a point grid and a volumetric grid.
 The volumetric grid determines the spatial regions.
 Each region ${\cal V}$ is actually a set of points that belong to the point grid.
 \item Volume shape:
 The spatial regions inherent to the V-SRP may follow arbitrary shapes and sizes, whereas the M-SRP
 assumes cubic regions surrounding each point of the grid.
 \item Lag weights:
 V-SRP includes the weights $\chi_p[\zeta,{\cal V}]$, which allow one to skip all values of $\phi_p[\zeta]$ not
 {associated with the adopted point grid} and, as a consequence, providing a desirable control over the {\it trade-off between
 computational complexity and accuracy}.
 \item TDoA bounds:
 In the V-SRP, the lags $\zeta_{p,{\cal V}}^{\rm min}$ and $\zeta_{p,{\cal V}}^{\rm max}$ are pre-computed by checking
 the TDoAs related to all grid points inside the volume ${\cal V}$.
 In the M-SRP, however, the TDoA bounds ${\hat\zeta_{p,{\bf x}}^{{\rm M}, {\rm min}}}$ and
 ${\hat\zeta_{p,{\bf x}}^{{\rm M}, {\rm max}}}$ are coarse estimates of the minimum and maximum TDoAs
 inside a cube surrounding $\mathbf{x}$ for the $p$th microphone pair.
\end{enumerate}

In order to numerically exemplify the difference between V-SRP and M-SRP, consider item 4 above. Due to the approximation adopted by M-SRP, it is easy to find examples in which the set of lags
$\{ {\hat\zeta_{p,{\bf x}}^{{\rm M}, {\rm min}}}, \ldots, {\hat\zeta_{p,{\bf x}}^{{\rm M}, {\rm max}}} \}$
either includes TDoAs not found or does not include all TDoAs found within the volume,
as in the next example. Let the sampling rate be $f_s = 48$~kHz, the speed of sound be 340~m/s, the two microphones of a given $p$th microphone pair be
located at ${\bf m}_{p,1} = [-2 \ 0 \ 0]^T$~m and ${\bf m}_{p,2} = [2 \ 0 \ 0]^T$~m, and the center of the cube with $r = 1$-m length edges be
located at $[0 \ 2 \ 0]^T$~m. {Following the equations in Section~\ref{sec:cobos},}
one arrives at $\hat\zeta_{p,{\bf x}}^{{\rm M}, {\rm min}} = -100$ and $\hat\zeta_{p,{\bf x}}^{{\rm M}, {\rm max}} = 100$.
However, by directly checking the TDoAs corresponding to the vertices of the cube, the minimum and maximum lags are found as
respectively $-110$ and $110$. Clearly, in this example several TDoAs associated with points within
the cube would be left out. Such issue might prevent the M-SRP from localizing sources placed close to the borders of the cube.

\section{Remarks on Implementation}\label{sec:implementation}

In order to maximally reduce the overall number of arithmetic operations required by each method in real time, and to enable a fair
comparison of their computational costs, the strategy of {\it pre-computing whatever is possible} is
assumed in this paper.
Such strategy yields a significant reduction of the computational burden in the search stage at the expense of requiring a larger
memory to store look-up tables.

Observe that SRP-based methods basically compute three quantities: (i) TDoAs, (ii) cross-correlations,
and (iii) {objective function} values for each grid element.
The implementation of each of these computations is described in the next subsections.

\subsection{Computing the TDoAs}\label{sub:comp_tdoa}

Prior to any processing and as soon as the spatial grid is defined, the TDoAs can be pre-computed and stored in look-up tables.
For the C-SRP method, given the position of the microphones and the points of the grid, all TDoAs
${\zeta}_p({\bf x})$ can be computed for all microphone pairs.
Thus, $P$ look-up tables are constructed, in which each entry is indexed by a point of the grid and stores its corresponding TDoA.

As for the V-SRP method, a table whose entries are indexed by volumes can be constructed as well.
In this case, each entry stores the set ${\cal Z}_{p,{\cal V}}$,
which corresponds to a list of lags $\zeta$ where $\chi_p[\zeta,{\cal V}]=1$ (see Eq.~\eqref{eq:vsrp_chi}), i.e.
\begin{align}\label{eq:Z_pV}
 {\cal Z}_{p,{\cal V}} \triangleq \left\{\zeta \in\mathbb{Z} \;\left|\; \chi_p[\zeta,{\cal V}]=1\right.\right\}.
\end{align}
By doing so, the expression actually implemented is
\begin{align}\label{eq:vsrp_cost_func_actually}
 \overline{W}({\cal V})
= \sum_{p=1}^P \sum_{\zeta \in  {\cal Z}_{p,{\cal V}}} \phi_p[\zeta],
\end{align}
whose main advantage over Eq.~\eqref{eq:vsrp_cost_func} is skipping trivial multiplications by $1$ and $0$.

In this subsection we presented an initialization procedure applicable to all SRP-based methods.
Since the TDoAs do not vary with time, for a fixed grid and array, they can be computed just once (during initialization) and
stored in look-up tables.
This strategy can significantly reduce the number of arithmetic operations performed in the
long run, especially when one is dealing with large rooms and/or using dense grids.

\subsection{Computing the Cross-Correlation Function}\label{sec:comp_cost}

The cross-correlation function {(CCF)} $\phi_p$ defined in Eq.~\eqref{eq:phi_p} can be pre-computed as soon as each
signal frame reaches the microphones. Aiming at real-time applications, it is preferable to compute $\phi_p$
in the frequency domain, using a fast Fourier transform (FFT) algorithm.
Besides reducing the number of {arithmetic} operations, working on the frequency domain is amenable to the use of PHAT.

It should be noticed that all SRP-based methods considered in this paper require the computation of $P$
CCFs and, therefore, the number of arithmetic operations required at this step is the same for these methods.

\subsection{Objective Function Evaluation}\label{sub:comp_funcEval}

For the C-SRP method, since the values of ${\zeta}_p({\bf x})$ are already stored in look-up tables,
the evaluation of the {objective function} given by Eq.~\eqref{eq:srp_cost_func}
involves only $P-1$ additions per point of the grid.
As for the V-SRP method, provided that $\chi_p[\zeta,{\cal V}]$ is efficiently stored, as explained in Subsection~\ref{sub:comp_tdoa}, the
evaluation of the {objective function} given by Eq.~\eqref{eq:vsrp_cost_func_actually} for a given volume ${\cal V}$
requires only $\left( \sum\limits_{p=1}^P  |{\cal Z}_{p,{\cal V}}| \right) -1$
additions, where $|{\cal Z}_{p,{\cal V}}|$ is the cardinality of the set ${\cal Z}_{p,{\cal V}}$ defined in Eq.~\eqref{eq:Z_pV}.
In the next section, the number of computations performed in this step is detailed for both the C-SRP and the V-SRP.



\section{Number of Arithmetic Operations per Frame}\label{sec:num_op}

It is known that the number of arithmetic operations required by {objective function} evaluations tend to be dominant in the long
run~\cite{Benesty2007}, and
this section addresses this issue.
{Observe that, as explained in Subsection~\ref{sub:comp_funcEval}, only one type of arithmetic operation is
actually required: summation.}
Specifically, given the size of the search space and the definition of a grid,
approximations for the {\it number of arithmetic operations per frame due to {objective function} evaluations} are provided
for the C-SRP and V-SRP methods.

\begin{figure}[t!]
\begin{center}
\includegraphics[width=0.9\linewidth]{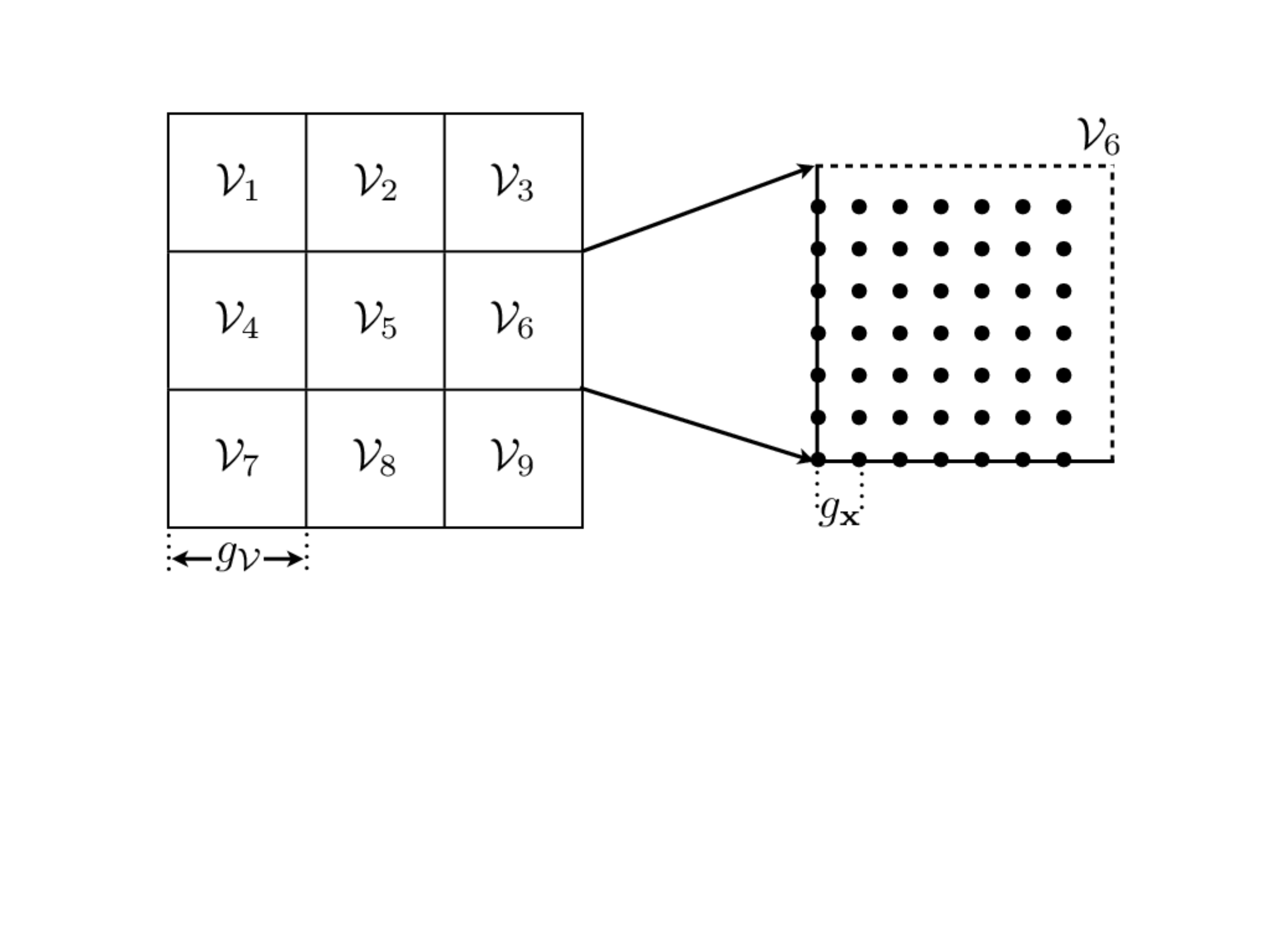}\vspace*{-1.9cm}
\caption{Example of a 2-D volumetric grid: the volumes degenerate to {planar} regions.} \vspace{-.3cm}
\label{fig:vols}
\end{center}
\end{figure}

Assume that the search space has the shape of a rectangular parallelepiped with length $L$, width $W$, and height $H$, all of them being positive real numbers.
In addition, let $g_{\bf x}\in\mathbb{R}_+$ denote the smallest distance between adjacent points of the grid and
$g_{\cal V}\in\mathbb{R}_+$ denote the smallest distance between adjacent volumes of the grid, i.e. each volume
is actually a set corresponding to the points of the grid within a cube of edge $g_{\cal V}$.
An illustration of a 2-D volumetric grid where the quantities $g_{\bf x}$ and $g_{\cal V}$ appear is
shown in Fig.~\ref{fig:vols}. Observe that these quantities represent a uniform sampling of the search space in terms
of points and volumes.

The number of points within the grid, $N_{\rm g}\in\mathbb{N}$, is given by
\begin{align}
 N_{\rm g}
 & = \left\lfloor\left( \frac{L+g_{\bf x}}{g_{\bf x}} \right) \left( \frac{W+g_{\bf x}}{g_{\bf x}} \right) \left( \frac{H+g_{\bf x}}{g_{\bf x}} \right)\right\rfloor
\end{align}
In addition, the number of volumes within the volumetric grid, $N_{\rm v}\in\mathbb{N}$, is
\begin{align}
 N_{\rm v}
 &=  \left\lfloor \frac{L W H}{g_{\cal V}^3} \right\rfloor
\end{align}

The number of points and volumes in a grid {determines} how many times the {objective function} is evaluated.
Therefore, the total number of arithmetic operations for the C-SRP method, $N_{\rm op}^{\textrm{C-SRP}}\in\mathbb{N}$,
{can be written as}
\begin{align}\label{eq:numOp_csrp}
 N_{\rm op}^{\textrm{C-SRP}} = N_{\rm g} \left( P-1 \right) .
\end{align}
On the other hand, for the V-SRP method, the total number of arithmetic operations, $N_{\rm op}^{\textrm{V-SRP}}\in\mathbb{N}$, is given by
\begin{align}
 N_{\rm op}^{\textrm{V-SRP}}
 &= \sum_{{\cal V} \in \Gamma} \left[ \left( \sum_{p=1}^P |{\cal Z}_{p,{\cal V}}| \right) -1 \right],   \label{eq:numOp_vsrp}
\end{align}
where $\Gamma$ is a set containing all volumes. Note that, by defining the average cardinality of the set ${\cal Z}_{p,{\cal V}}$ as
\begin{align}
 \langle|{\cal Z}|\rangle = \frac{1}{N_{\rm v}P}\sum_{{\cal V} \in \Gamma}\sum_{p=1}^P|{\cal Z}_{p,{\cal V}}|,
\end{align}
then Eq.~\eqref{eq:numOp_vsrp} can be rewritten as
\begin{align}
 N_{\rm op}^{\textrm{V-SRP}}
 &= N_{\rm v} \left( P \langle|{\cal Z}|\rangle -1\right) \label{eq:numOp_vsrp2}
%
\end{align}

Equations~\eqref{eq:numOp_csrp} and~\eqref{eq:numOp_vsrp2} represent the number of arithmetic operations per frame
due to objective function evaluations for the C-SRP and V-SRP, respectively.
Observe that we are not taking into account the computations of the TDoAs and CCFs due to the reasons explained in
Section~\ref{sec:implementation}.
In order to fully understand these expressions,
consider that the edge of each cube and the distance between adjacent points of the grid are related by
$g_{\cal V} = \alpha g_{\bf x}$, where $1 < \alpha \in \mathbb{N}$.
In this case, $N_{\rm op}^{\textrm{V-SRP}}$ and $N_{\rm op}^{\textrm{C-SRP}}$ {are} related by
\begin{align}
 N_{\rm op}^{\textrm{V-SRP}}
 &= \left\lfloor \frac{L W H}{g_{\cal V}^3} \right\rfloor ({P  \langle|{\cal Z}|\rangle} - 1)
 \nonumber\\
 &< \frac{L W H}{\alpha^3 g_{\rm x}^3} {P  \langle|{\cal Z}|\rangle}
 \nonumber\\
 &< \frac{(L+g_{\rm x}) (W+g_{\rm x}) (H+g_{\rm x})}{g_{\rm x}^3} P  \frac{\langle|{\cal Z}|\rangle}{\alpha^3}
  \approx  N_{\rm op}^{\textrm{C-SRP}}\frac{\langle|{\cal Z}|\rangle}{\alpha^3} ,
 \label{eq:numOp_vsrp3}
\end{align}
where the approximation is valid as long as the number of microphone pairs $P \gg 1$, which usually is the case.
Additionally, one may regard $L, W, \text{ and } H$ as multiples of $g_{\rm x}$ in order to disregard the floor operator.

In addition, note that the number of points within a cube of edge $g_{\cal V} = \alpha g_{\bf x}$ is
$\alpha^3$.
Therefore, in the {\it worst case scenario}, each point of the cube leads to a different TDoA implying that
the maximum number of  different TDoAs in a volume is $\alpha^3$ (for a given microphone pair).
Thus, one has $\langle|{\cal Z}|\rangle \leq \alpha^3$, in which the equality is achieved only when all
volumes fall into the {\it worst case scenario} for all microphone pairs, a phenomenon {that has not been observed with the data we tested}.
Therefore, the computational cost of the V-SRP method is, in the worst case, equivalent to the one of the C-SRP.
However, it was observed that  many points within a volume lead to the same TDoA and, consequently,
the number of arithmetic operations per frame of the V-SRP is usually much lower than the one of the C-SRP method, since
it is rather common to have $\langle|{\cal Z}|\rangle \ll \alpha^3$, especially for volumes relatively far from the array,
as it was illustrated by the discriminability index results presented in~\cite{Lonnes_iwaenc2012}.
{Moreover, the minimum value for $\langle|{\cal Z}|\rangle$ is $1$, at least theoretically.
In such case, the V-SRP would perform about $\alpha^3$ {times} fewer arithmetic operations, as compared to the C-SRP.}

For the RV-SRP method, the number of {arithmetic} operations per frame can be determined by summing the following
two terms: (i) number of {arithmetic} operations for the V-SRP considering the entire search space and
(ii) number of {arithmetic} operations required by the C-SRP considering that the size of the search space is
reduced to the size of the winning volume.
If one compares the costs of the C-SRP and V-SRP, it is possible to see the motivation behind the development of {the} RV-SRP.
By using a coarse {volumetric grid}, it is possible to reduce the computational complexity of the V-SRP.
On the other hand, by applying the C-SRP to a smaller search region, its number of {arithmetic} operations are
drastically reduced (smaller $L$, $W$, and $H$ values), allowing the use of a denser {internal} grid
{in the second stage.}\footnote{{Under such conditions, previous computation of the TDoAs required by the refining stage could be avoided, since its contribution to the overall complexity of the method is marginal.}} As will be shown in the next section, by choosing the grids appropriately, one can achieve a low estimation error with low computational complexity when using the RV-SRP method. {Of course, if the volumetric grid becomes too coarse, the overall complexity of the RV-SRP method tends to increase again, since the high cost of the C-SRP refining stage applied to large volumes dominates.}


\section{Results}\label{sec:results}

The performance of the proposed methods have been assessed through experiments with simulated and recorded signals,
as explained in this section. The target here is to point out some attractive features of the proposed
algorithms, as well as to highlight some trade-offs between computational complexity and localization performance.
The C-SRP and M-SRP methods are used as benchmarks for comparisons.\footnote{The PHAT pre-filtering is employed when computing the related cross-correlations in all methods.}
Two experiments are described: one using acquired signals (Subsection~\ref{sub:exper}) and the other using
simulated signals (Subsection~\ref{sub:artif}).
{After having investigated the performance of the proposed methods, a brief discussion on how to set
the grids is provided (Subsection~\ref{sub:rel_grids}).}

\subsection{Data from a Real Scenario}\label{sub:exper}


In this subsection, the performances of both V-SRP and RV-SRP methods are assessed when applied to signals acquired by a uniform linear array (ULA) in a reverberant room. First the experimental setup is described, then the results are presented along with a brief discussion of the trade-offs between computational complexity and accuracy.

\subsubsection{Experimental Setup}

The experiments are conducted in a $5.2\;{\rm m}\times 7.5\;{\rm m}\times 2.6\;{\rm m}$  room whose measured T60 is approximately $500$~ms.
The microphone array is a ULA composed of $8$ microphones and with aperture of $2.1$~m.
A small-size loudspeaker ($10$-cm diameter)\footnote{{Since} {the loudspeaker is not a point source, there is an inherent uncertainty relative to the source position which limits the minimum error that can be achieved by the methods.}} plays the role of the single acoustic source.
The source signal consists of $3$ sentences emitted by a female speaker and has a total duration of $4.5$~s.
The sentences were recorded in a professional studio and PCM-coded with a sampling rate of $48$~kHz and $24$-bit precision.
A voice-activity detector (VAD) is employed before playing back the signals in order to discard speech-free segments of the original source signal.

\begin{figure}[!t]
\centering
\includegraphics[width=0.55\linewidth]{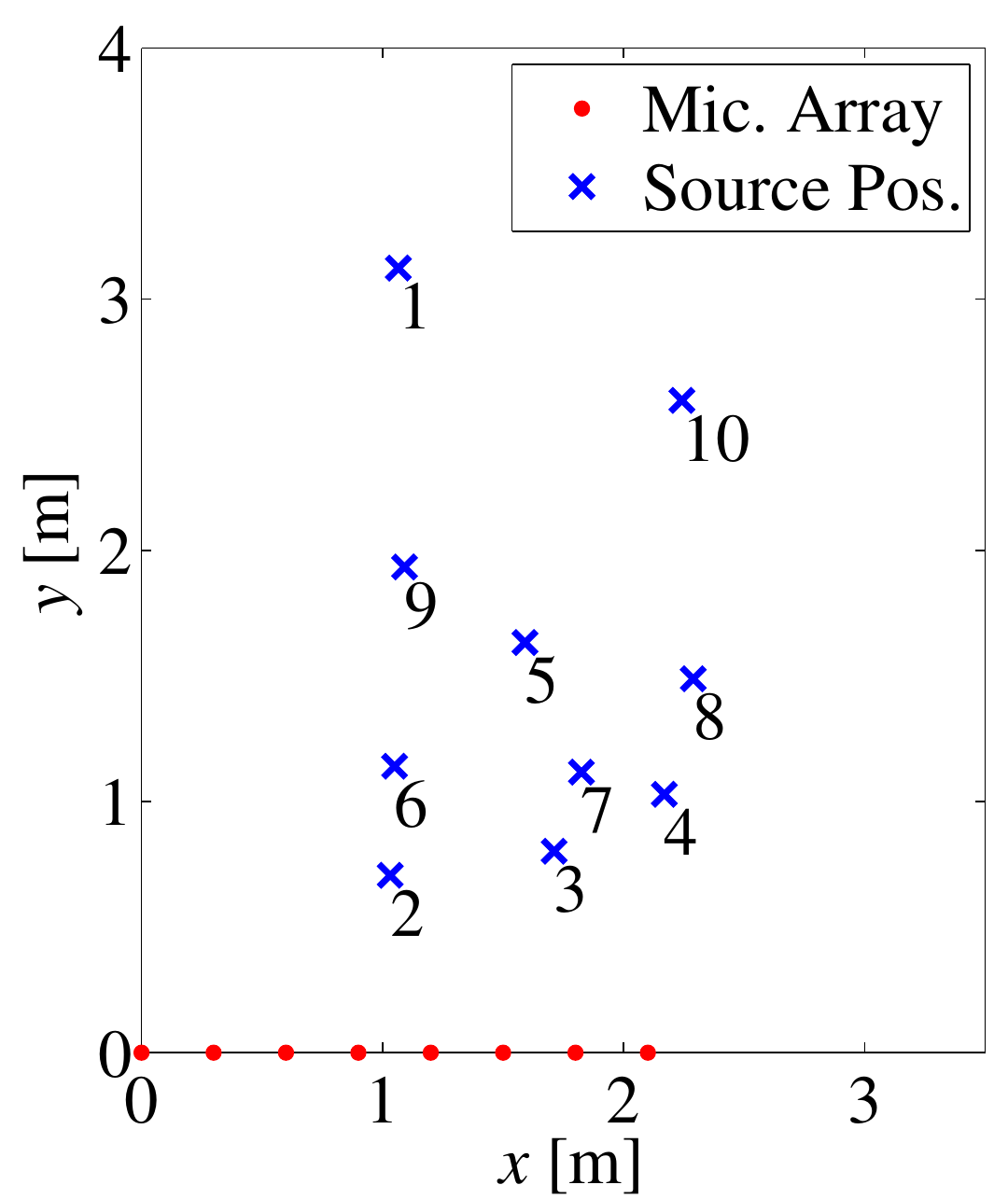}
\caption{Positions of microphones and sound sources (Subsection~\ref{sub:exper}).}
\label{fig:sourcePositions}
\end{figure}

The loudspeaker is placed at $10$ different positions chosen at random, as illustrated in Fig.~\ref{fig:sourcePositions}.
Both microphones and source are always at the height of $72.5$~cm.
The sound source is amplified at the loudspeaker output in such a way to maximize the signal-to-interference-plus-noise ratio (SINR) at the microphones without saturating any of the amplifiers on the signal path.

\subsubsection{Setup of the Localization Methods}

Due to the inherent limitations associated with the ULA geometry to localize sources in the 3-D Euclidean space,
and as the sources and microphones are always at the same height, the source location is estimated over the $xy$-plane whose height is $72.5$~cm.
The search region is the square with opposite vertices $(0, 0, 0.725)$~m and $(3.5, 4.0, 0.725)$~m, as illustrated in Fig.~\ref{fig:sourcePositions}.

All the source localization methods were applied in successive 4096-sample long frames { (85 ms at 48{-}kHz sampling rate)} with 50\% of overlap. Considering that the signals have a duration of 4.5 s and that there are 10 different source positions, then a total of 1040 positions were estimated by each method. The error of each estimate was calculated as the Euclidean distance between the actual and estimated source position considering only $x$ and $y$ coordinates.

\begin{figure*}[!t]
\centering
\subfigure[Grid resolution of $1$~cm.]{\includegraphics[scale=0.4]{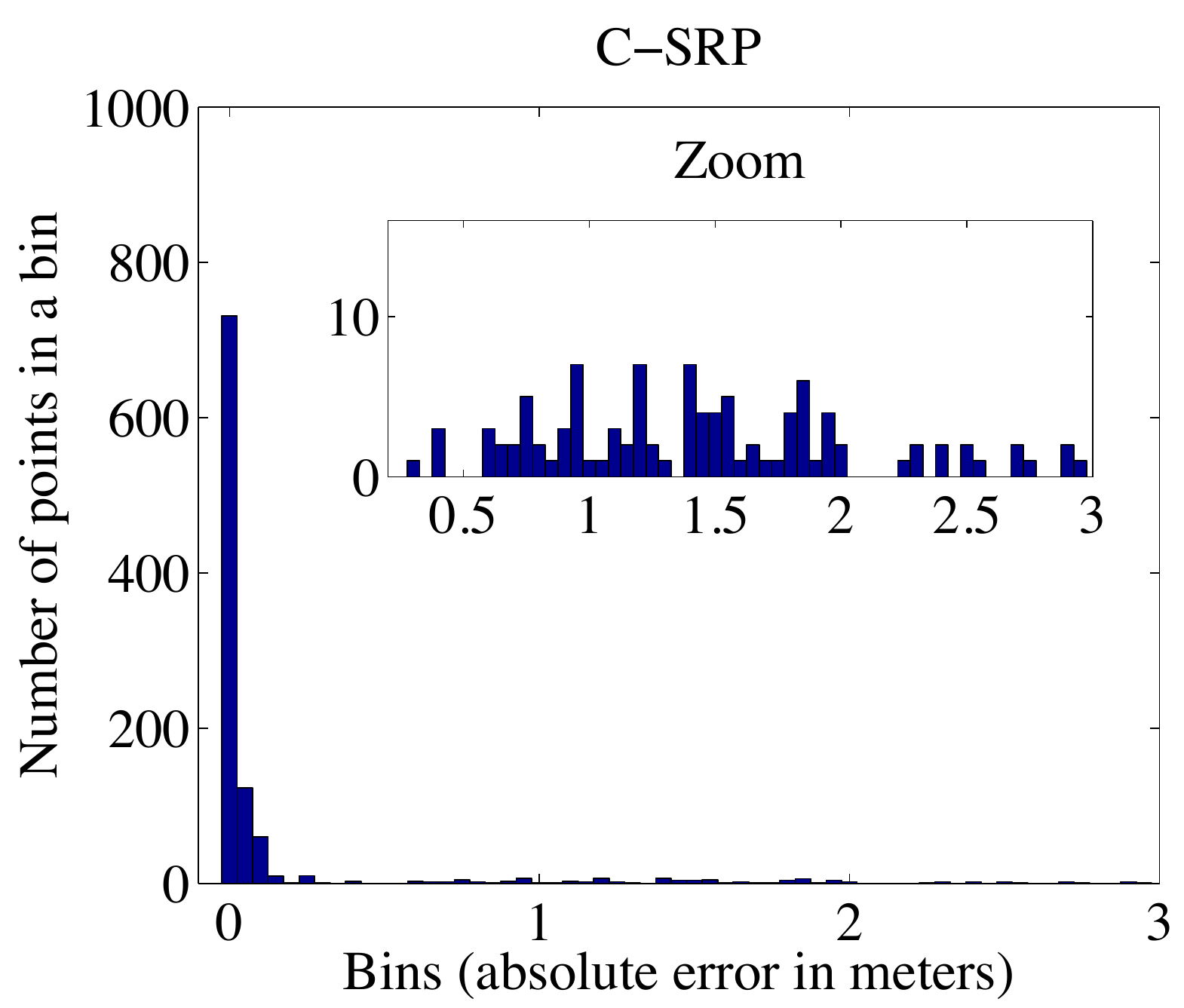}
\label{fig:1cmsrp_exper}}
\subfigure[Grid resolution of $10$~cm.]{\includegraphics[scale=0.4]{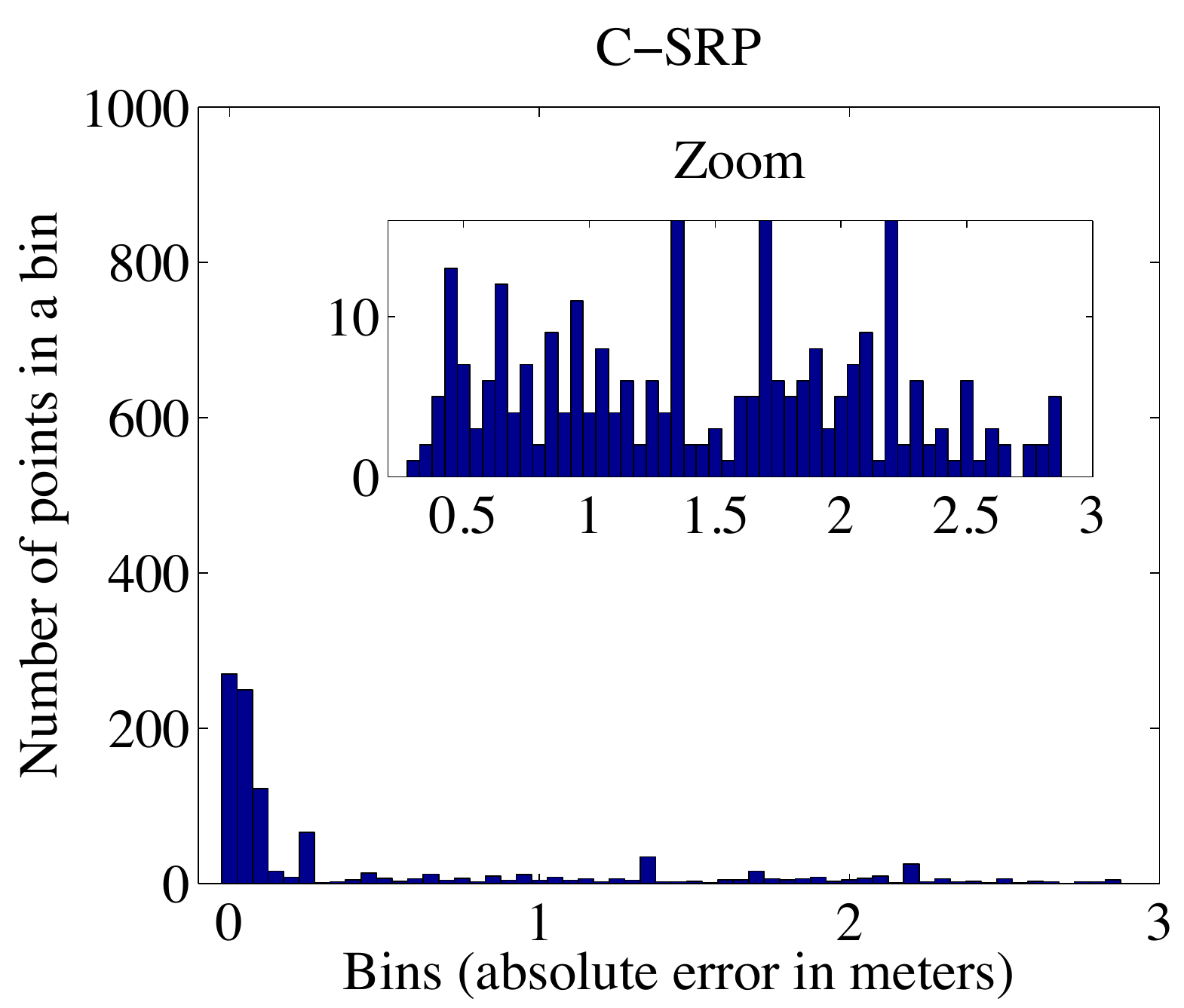}
\label{fig:10cmsrp_exper}}
\caption{Histograms of location estimation errors for the C-SRP (bin width is $5$~cm). {The inside histograms (Zoom) show the number of estimation errors larger than 30~cm.}\label{fig:srp_exper}}
\end{figure*}
\begin{figure*}[!t]
\centering
\subfigure[Grid resolution of $10$~cm.]{\includegraphics[scale=0.4]{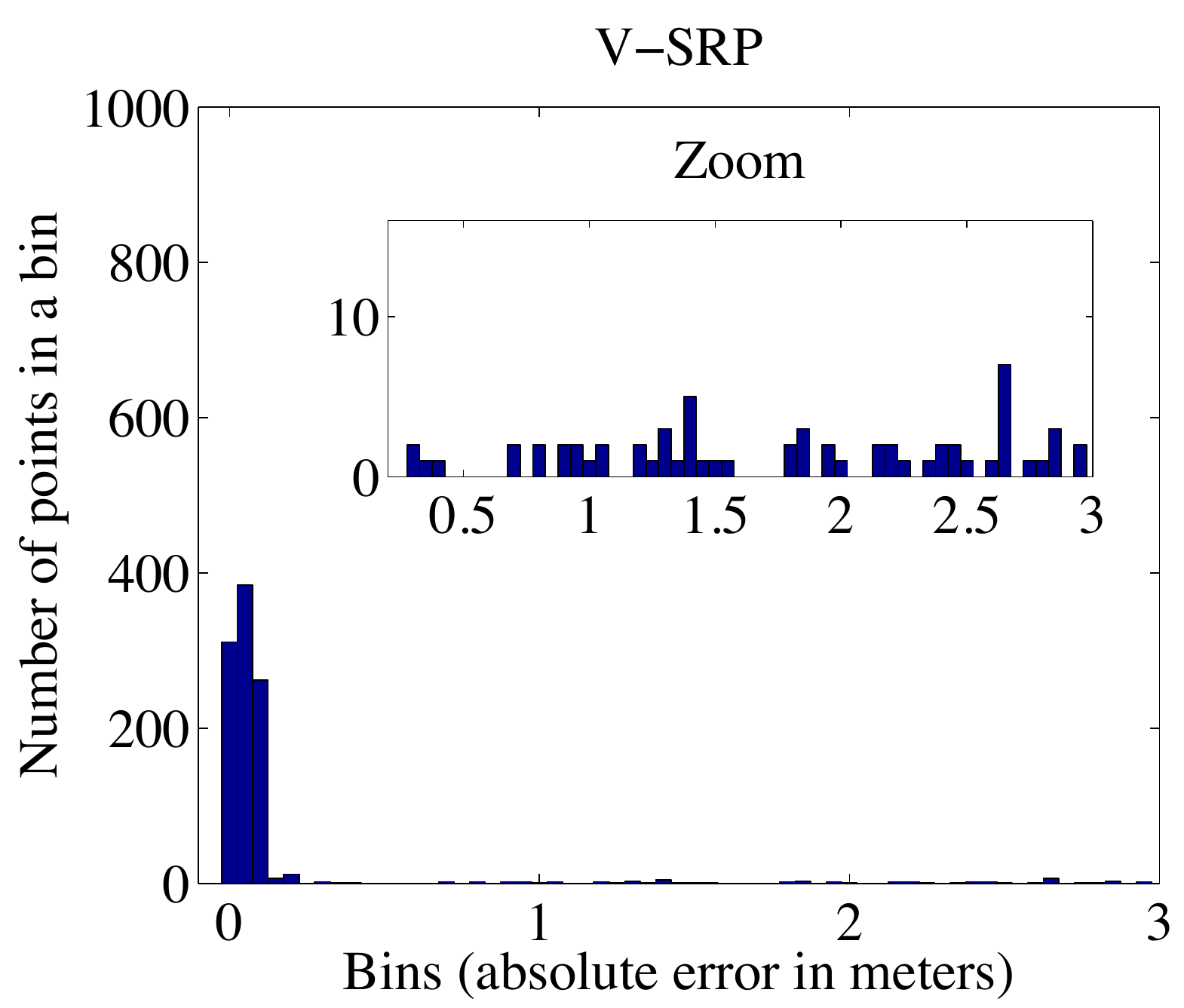}
\label{fig:10cmvolsrp_exper}}
\subfigure[Refinement of $1$~cm.]{\includegraphics[scale=0.4]{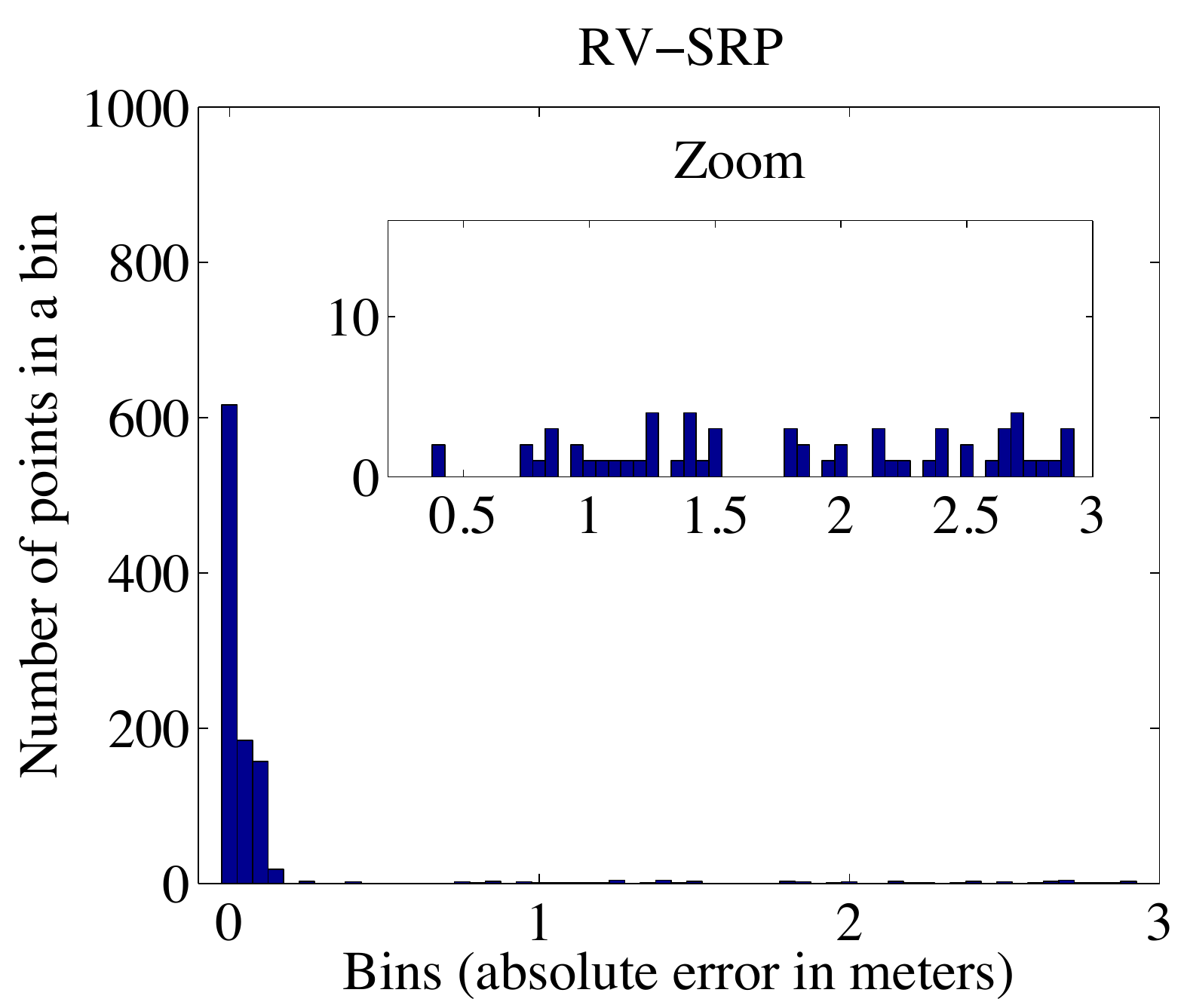}
\label{fig:10cmvolsrpref_exper}}
\caption{Histograms of location estimation errors for the V-SRP and the RV-SRP (bin width is $5$~cm). {The inside histograms (Zoom) show the number of estimation errors larger than 30~cm.}\label{fig:volsrp_exper}}
\end{figure*}

The C-SRP method was run with two different search grids, with 1-cm and 10-cm distance between adjacent points respectively. These two values will be used to illustrate the direct compromise between grid resolution and position estimation error. The V-SRP method was run over a grid of squares of 10-cm edges, each one enclosing 16 grid points (refer to Fig.~\ref{fig:vols}). Regarding the RV-SRP method, a refinement stage using a grid of 1-cm is employed on the winning volume of the V-SRP. The performance of the M-SRP method with points spaced 10-cm apart was also evaluated.

\subsubsection{Results and Discussion}

\begin{figure}
\centering
\includegraphics[scale=0.4]{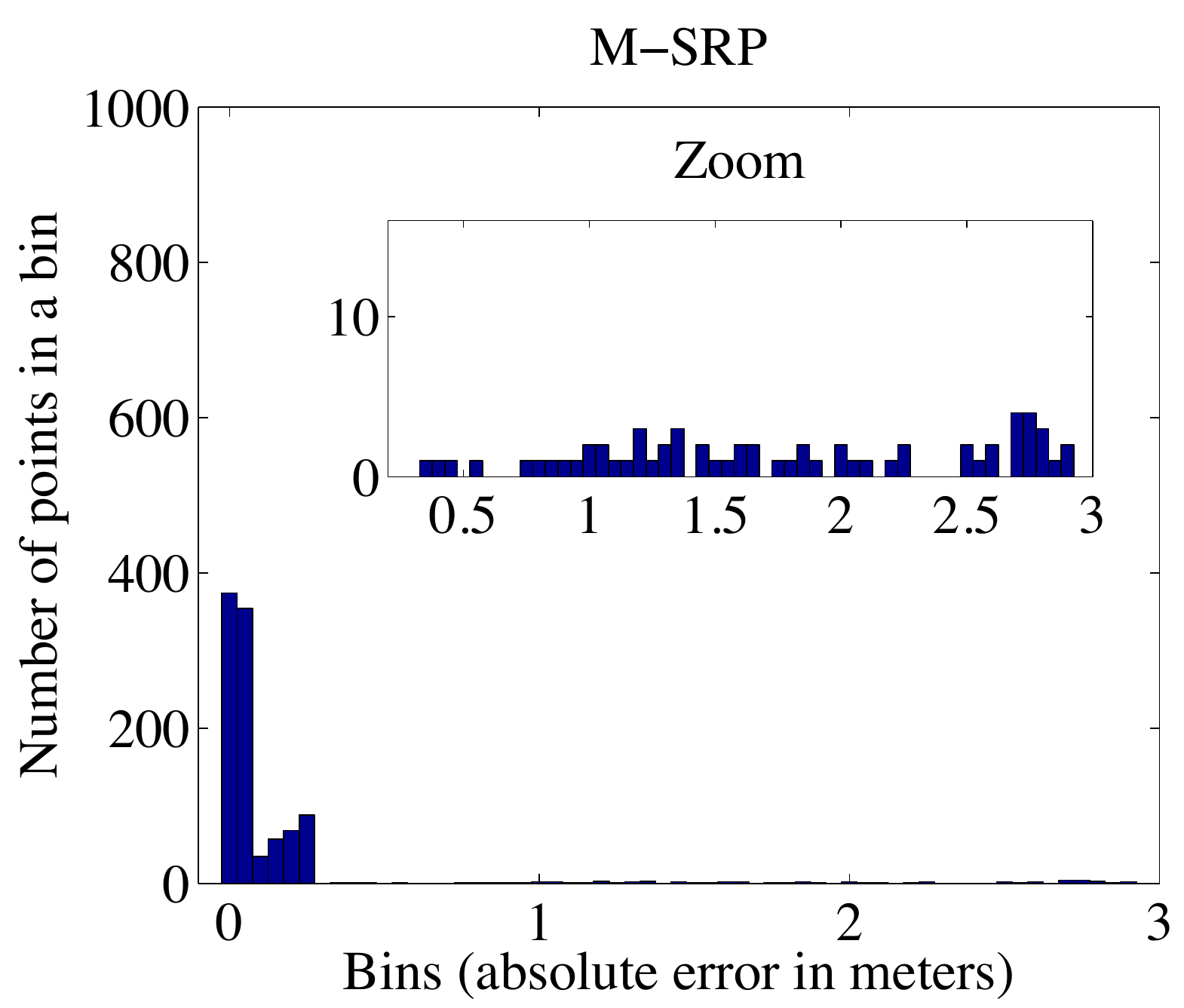}
\caption{Histograms of location estimation errors for the M-SRP (bin width is $5$~cm){. The inside histogram (Zoom) shows the number of estimation errors larger than 30~cm.}}
\label{fig:cobos_exper}
\end{figure}

Figs.~\ref{fig:srp_exper}, \ref{fig:volsrp_exper}, and~\ref{fig:cobos_exper} show histograms of the estimation errors of the C-SRP, the (R)V-SRP, and the M-SRP, respectively.
Those histograms consider all 1040 estimates for each method and for their different configurations.
Table~\ref{tab:real_data_summary} displays the mean {and median} errors for each method along with the associated number of
{arithmetic} operations performed during the search stage.

\begin{table*}[!t]
\caption{Summary of the results for the real-data scenario.}\label{tab:real_data_summary}
\centering
{
\begin{tabular}{|c|c|c|c|}
\hline
Method [grid resolution]        & Mean error [cm] & Median error [cm] & Approx. number of op. per frame ($\times 10^5$) \\
\hline
C-SRP  [$1$~cm]                 &  $19.62$  &  $3.15$  &  $38.0$  \\
C-SRP  [$10$~cm]                &  $52.09$  &  $11.17$  &  $0.398$  \\
V-SRP  [$10$~cm, $16$~pt]                &  $18.29$  &  $6.21$  &  $2.08$  \\
RV-SRP [$10$~cm, $16$~pt / ref. $1$~cm]  &  $15.98$  &  $3.77$  &  {$2.11$} \\
M-SRP  [$10$~cm]                &  $19.67$  &  $6.76$  &  $2.71$  \\
\hline
\end{tabular}
}
\end{table*}

One can verify that the majority of the location estimation errors of the C-SRP method with a 1-cm resolution are between $0$ and $5$~cm (Fig.~\ref{fig:1cmsrp_exper}), with a mean estimation error of {$19.62$}~cm {and a median estimation error of $3.15$~cm} (Table~\ref{tab:real_data_summary}).
In addition, it is also possible to see that there are very few frames whose associated C-SRP outputs give
completely wrong source position estimates, a very desirable characteristic.
But from a practical perspective, the method suffers from a striking drawback: the computational complexity
associated with the search stage may hinder the application of such a high-density grid.
In this experiment, for example, the  number of {arithmetic} operations due to the C-SRP {objective function} evaluation was
around $38.0\times10^5$ per $4096$-sample frame to perform a 2-D search.
This is really an issue that the designer of such a system must face.
The computational burden can dramatically increase as more microphones are added (quadratic dependence, see Eq.~\eqref{eq:numMicPairs}) and/or the number of grid points grows (as in the case of $3$-D regions with dense grids).

A possible solution to the high computational demands required by source-localization algorithms is to decrease the number of points within the grid search. When the $10$~cm resolution C-SRP is employed (see Fig.~\ref{fig:10cmsrp_exper}) to the same problem,
the number of {arithmetic} operations associated with the functional evaluations of the C-SRP {objective function} goes down to approximately $3.98\times 10^4$ per frame, thus decreasing around two orders of magnitude as compared to the $1$-cm resolution grid.
Nonetheless, the performance is dramatically sacrificed, since the number of anomalous estimates increases too much. Indeed, the mean value of the estimation error in this case is around {$52.09$}~cm{, whereas the corresponding median error is around $11.17$~cm}.

The V-SRP method (see Fig.~\ref{fig:10cmvolsrp_exper}) yielded mean {and median} estimation {errors} around {$18.29$~cm and $6.21$~cm, respectively},
outperforming the C-SRP with grid resolution of $10$~cm.
This mean value is even a bit smaller than the mean estimation error value for the C-SRP with $1$-cm resolution,
but when one compares Fig.~\ref{fig:1cmsrp_exper} with Fig.~\ref{fig:10cmvolsrp_exper}
it is straightforward to verify that most of the results of Fig.~\ref{fig:1cmsrp_exper} are better than the ones in Fig.~\ref{fig:10cmvolsrp_exper}{, a fact that is reflected in the median errors of those methods}.
The key advantage of the V-SRP method comes when one
compares the computational complexity of the C-SRP with $1$-cm resolution to the V-SRP with $10$-cm edges.
Indeed, the number of {arithmetic} operations associated with the functional evaluations in the search stage is around $2.08\times10^5$ per frame, which yields a significant reduction of the computational burden, without sacrificing performance significantly.

By employing the RV-SRP strategy, the total number of {arithmetic} operations is slightly increased to approximately
$2.11\times 10^5$ {arithmetic} operations per frame, while the mean {and median} estimation {errors are} reduced to {$15.98$~cm and $3.77$~cm, respectively}.
Such results let clear the inherent capability of the proposed RV-SRP method to trade off performance and computational
burden, thus providing an additional degree of freedom to the design of sound source localization systems.

The M-SRP method proposed in~\cite{Cobos_letter2011} is outperformed (see Fig.~\ref{fig:cobos_exper} and
Table~\ref{tab:real_data_summary}) in this particular experimental setup by both V-SRP and RV-SRP as regards
estimation error and number of {arithmetic} operations.

Finally, Table~\ref{tab:real_data_scenario} contains the actual number of {arithmetic} operations per frame required to
evaluate the {objective function}s over the entire search region for five different search grids.
In the cases of the V-SRP and RV-SRP, the resolution indicates the size of the edges of the square spatial regions, each one enclosing $16$ grid points.
In addition, the refinement of the RV-SRP is always implemented  by employing a C-SRP with a $1$-cm grid resolution within the {selected} volume.

\begin{table}[!t]
\caption{Number of {arithmetic} operations per frame due to functional evaluations. {(* means that the
RV-SRP coincides with the V-SRP for this resolution, i.e., no refinement is actually performed)}}\label{tab:real_data_scenario}
\centering
\begin{tabular}{|c|c|c|c|c|}
\hline
Resolution & C-SRP  &    V-SRP    &    RV-SRP & M-SRP  \\
\hline
$1$~cm   &  $3,800,277$ &  $5,821,039$  & {$5,821,039$*} &  $6,154,534$   \\
$10$~cm  &  $39,852$    &  $208,378$    & $211,078$    &  $270,682$   \\
$20$~cm  &  $10,206$    &  $79,419$     & $90,219$     &  $128,072$   \\
$50$~cm  &  $1,944$     &  $21,439$     & $88,939$     &  $53,932$   \\
\hline
\end{tabular}
\end{table}

\subsection{Simulated Scenario}\label{sub:artif}

In this subsection, the performance of the proposed algorithms is evaluated using simulated signals.  The objective of the simulation is to search for a very high localization accuracy in the $3$-D space, yet focusing on saving computational resources, especially as regards to decreasing
the amount of {arithmetic}  operations related to functional evaluations. In order to achieve such high accuracy results, one needs to properly choose the array geometry, the grid resolution, and the specific sound source localization method. The array geometry that allows for high $3$-D localization accuracy must have both relatively large aperture and great amount of microphone pairs, so that the resulting spatial resolution is substantially increased. Along with these choices, the spatial grid must also be dense enough. In the following, the simulation setup and the chosen localization methods are described along with their associated results.

\subsubsection{Simulation Setup}

The environmental setup simulated for this example consists of a $4.0~{\rm m}\times6.0~{\rm m}\times3.0~{\rm m}$ reverberant room
whose T60 {can be either $250$~ms or}  $500$~ms.
Such reverberant {environments are} simulated using the image model method~\cite{Allen1979_ImageMethod,Benesty2007}.
The speech signal employed is a $1$-s segment from the same source signal used in Subsection~\ref{sub:exper}.

\begin{figure}[!t]
\centering
\includegraphics[scale=0.45]{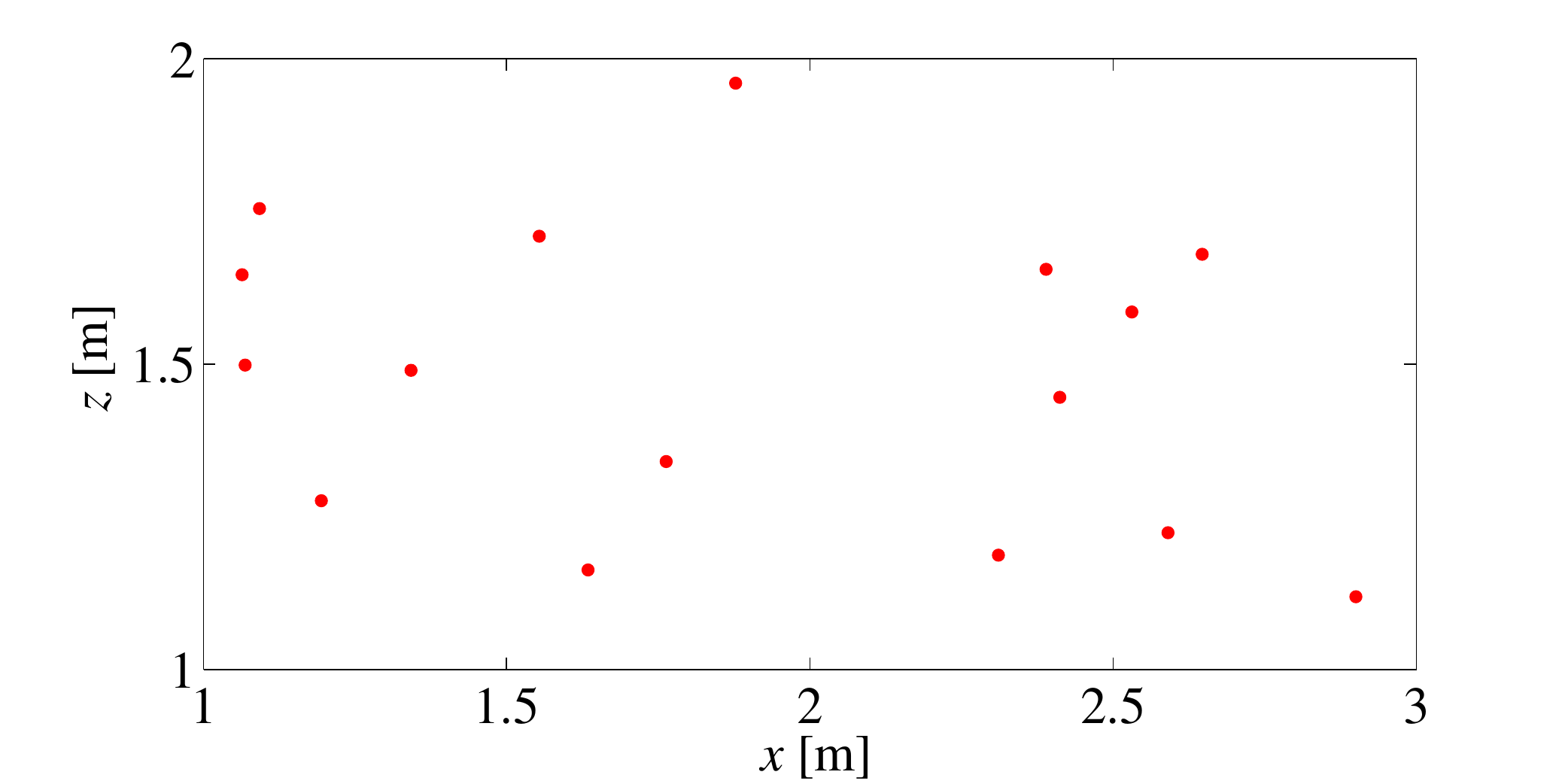}
\caption{Locations of each microphone in the planar array (Subsection~\ref{sub:artif}).}
\label{fig:planar_array}
\end{figure}

The array has 16 microphones distributed as depicted in Fig.~\ref{fig:planar_array} on a $2.0~{\rm m}\times1.0~{\rm m}$ region perpendicular to the floor plane (i.e. to the $xy$-plane), mounted on one of the walls of the room (at $y=0$). The source signal was artificially located at 5 different positions, which are shown in Table~\ref{tab:src_pos_artificial}.

\begin{table}[!t]
\caption{Source positions of the Artificial Scenario.}\label{tab:src_pos_artificial}
\centering
\begin{tabular}{|c|c|}
\hline
\# & Source Position \\\hline
1&[2.92, 2.18, 1.64] \\\hline
2&[2.09, 1.01, 1.88]\\\hline
3&[1.28, 2.13, 1.88]\\\hline
4&[1.30, 0.99, 1.69]\\\hline
5&[1.52, 2.36, 1.78]\\
\hline
\end{tabular}
\end{table}

\subsubsection{Setup of the Localization Methods}

In this scenario, the search space is the entire room (a $4.0~{\rm m}\times6.0~{\rm m}\times3.0~{\rm m}$) region, and the error is the 3-D Euclidean distance between estimated and known source positions. The C-SRP method employed a grid resolution of 3~cm.\footnote{As will be mentioned, the computational cost of the C-SRP using a 1-cm resolution over the whole 3-D space is too high for this experiment, {hence a coarser resolution is used.}} Since the objective of this experiment is to attain high accuracy, the RV-SRP was employed. The V-SRP step was run over cubic spatial regions with 10~cm edges enclosing $64$ points each. The refinement stage considered a search grid with 1-cm resolution. As a benchmark, the M-SRP with a resolution of 10 cm was also evaluated. As in the previous experiment, source position estimates were calculated for successive 4096-sample long frames with 50\% of overlap.

\subsubsection{Results and Discussion}

\begin{figure*}[!t]
\centering
\subfigure[{Grid res. of $3$~cm: T60~$ = 250$~ms.}]{\includegraphics[scale=0.35]{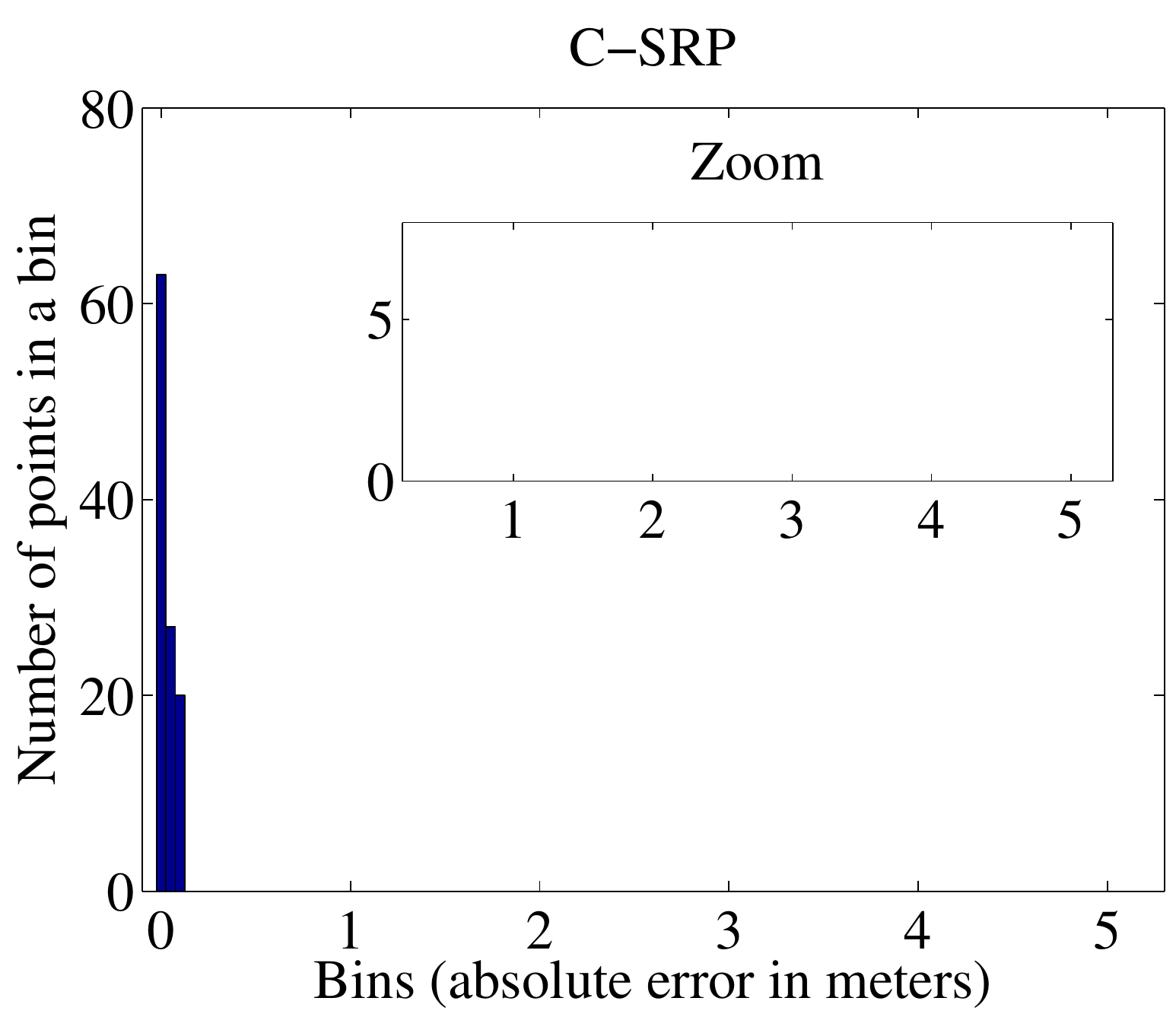}
\label{fig:3cmsrp_simu_250}}
\subfigure[{Grid res. of $10$~cm / ref. $1$~cm: T60~$ = 250$~ms.}]{\includegraphics[scale=0.35]{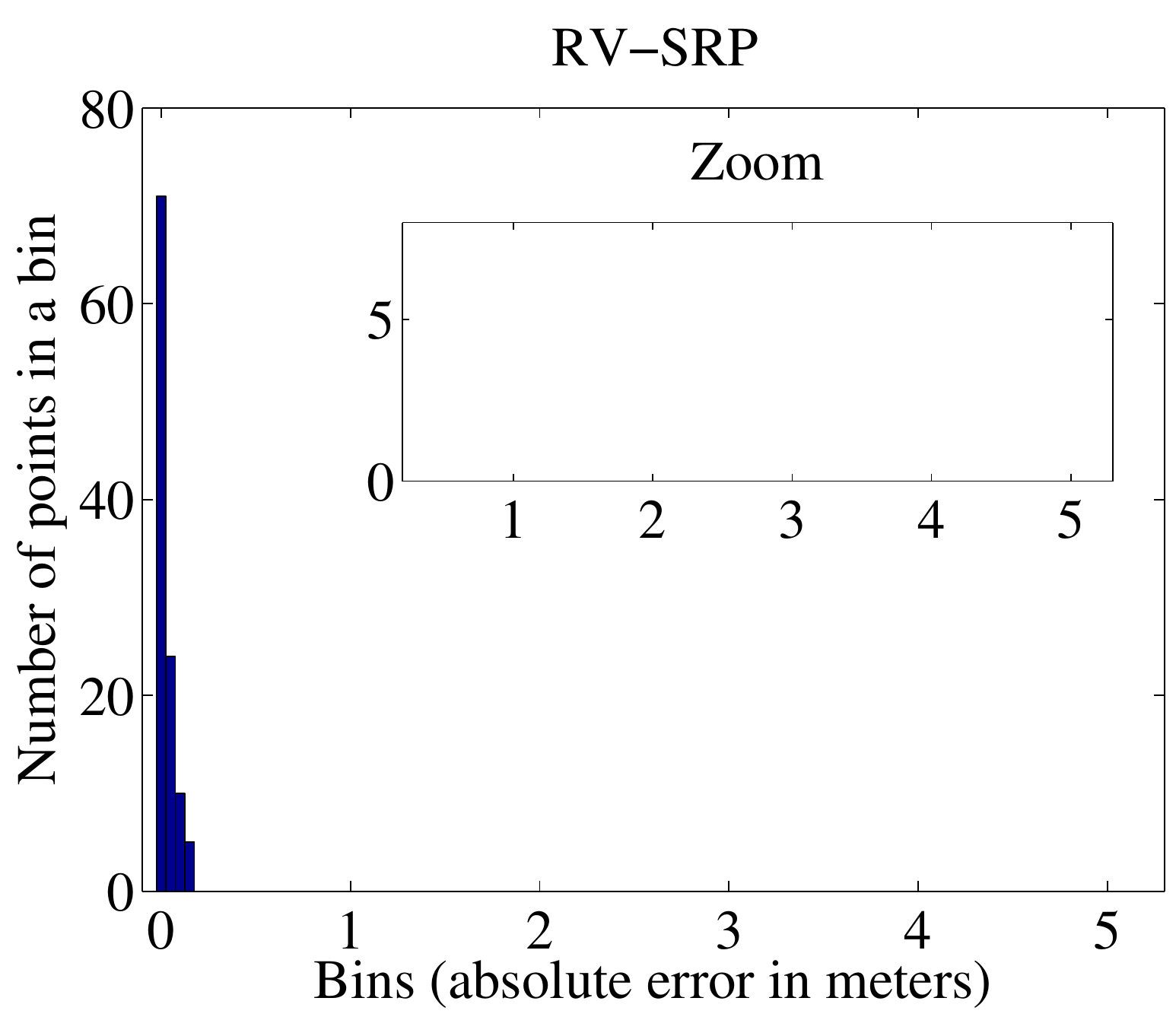}
\label{fig:10cmvsrp_simu_250}}
\subfigure[{Grid res. of $10$~cm: T60~$ = 250$~ms.}]{\includegraphics[scale=0.35]{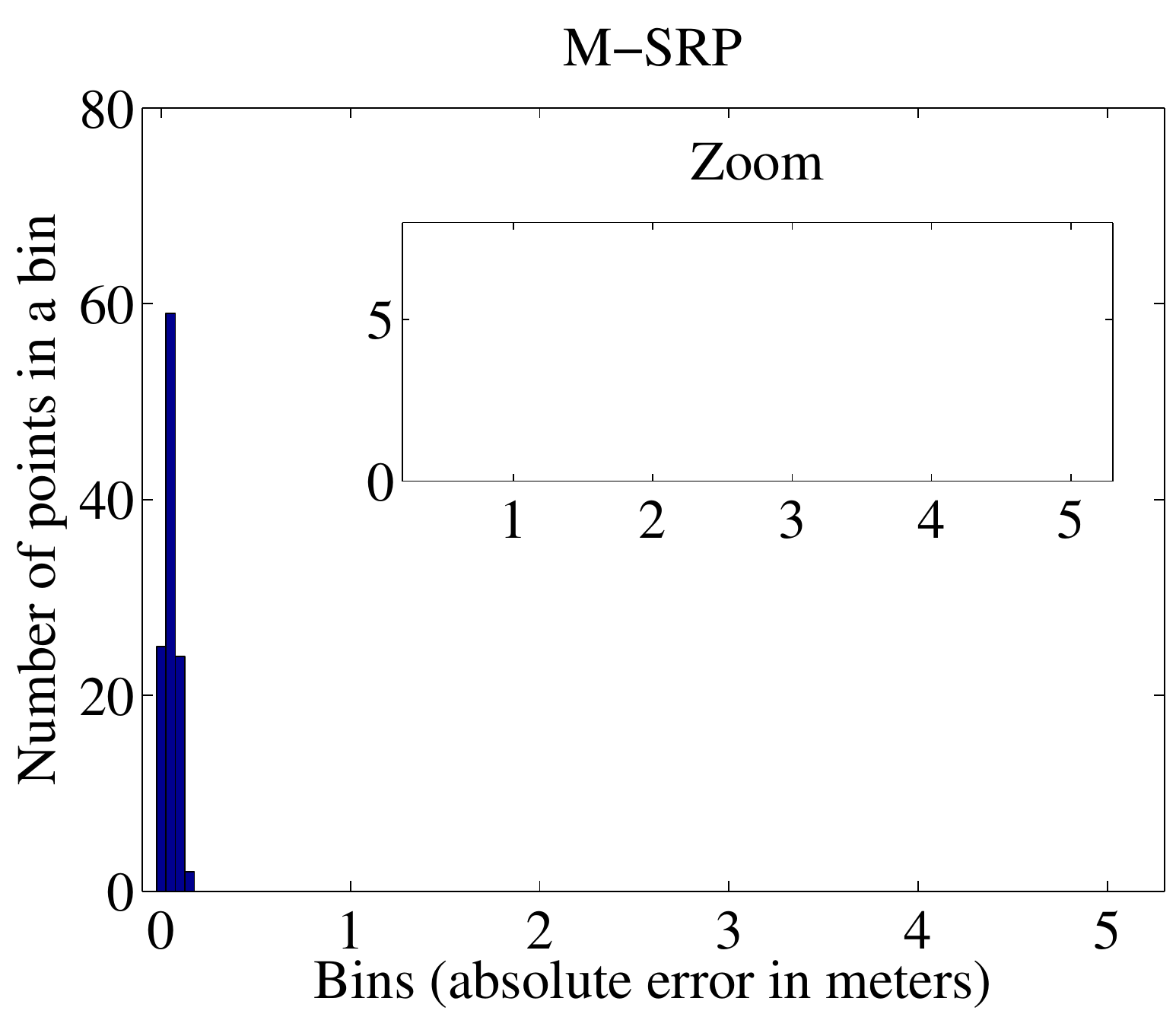}
\label{fig:cobos_simu_250}}\\
\subfigure[{Grid res. of $3$~cm: T60~$ = 500$~ms.}]{\includegraphics[scale=0.35]{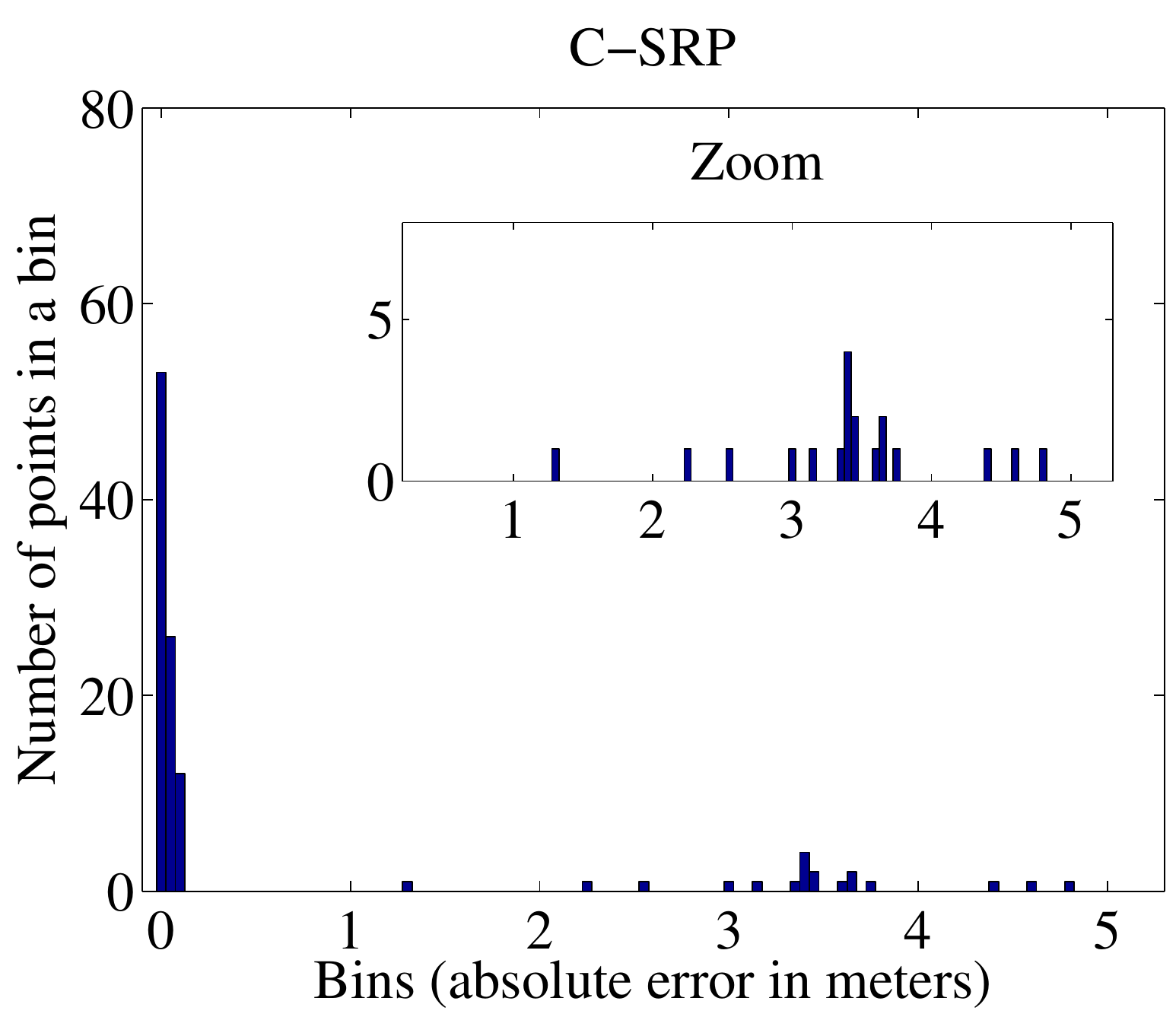}
\label{fig:3cmsrp_simu}}
\subfigure[{Grid res. of $10$~cm / ref. $1$~cm: T60~$ = 500$~ms.}]{\includegraphics[scale=0.35]{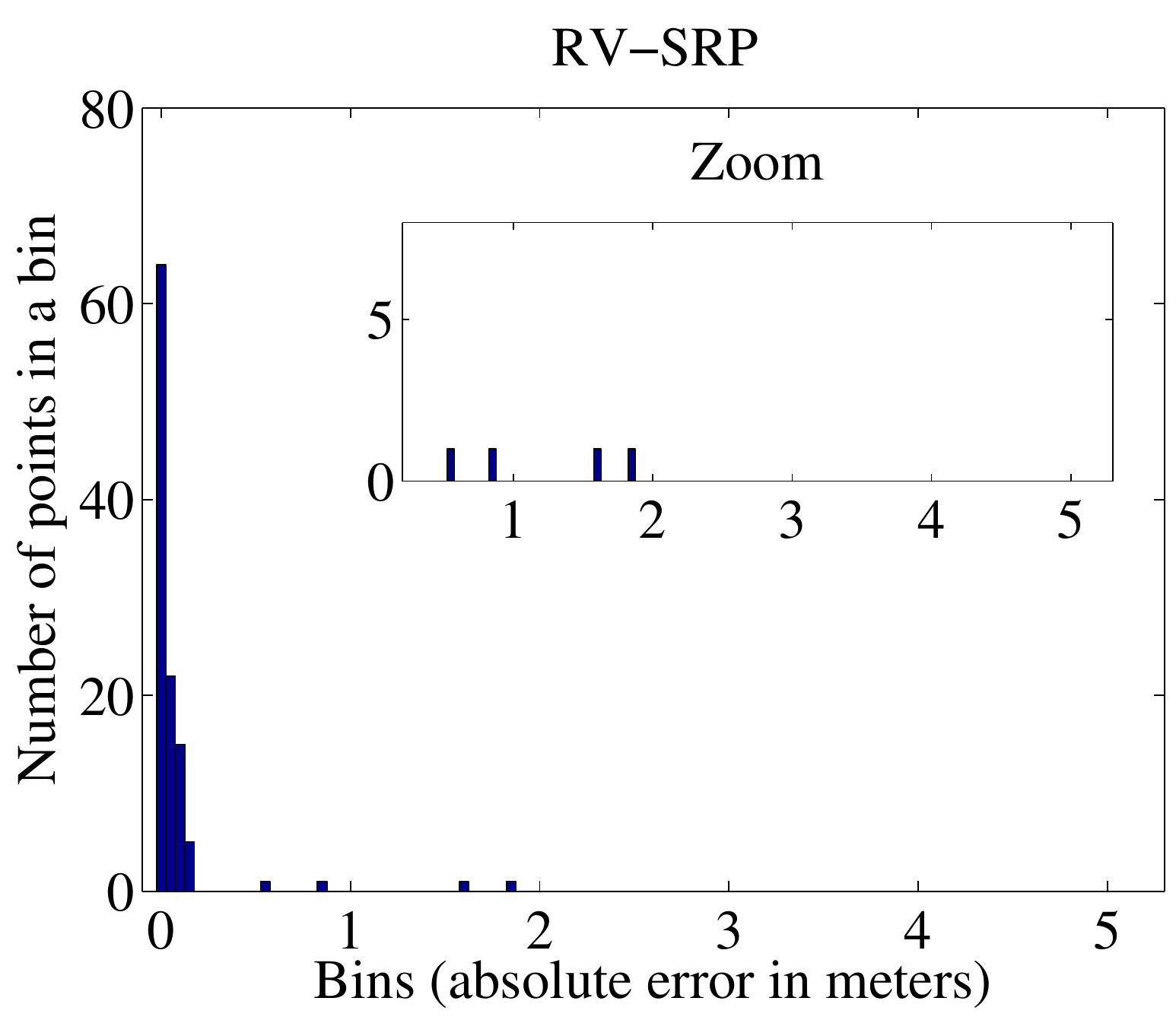}
\label{fig:10cmvsrp_simu}}
\subfigure[{Grid res. of $10$~cm: T60~$ = 500$~ms.}]{\includegraphics[scale=0.35]{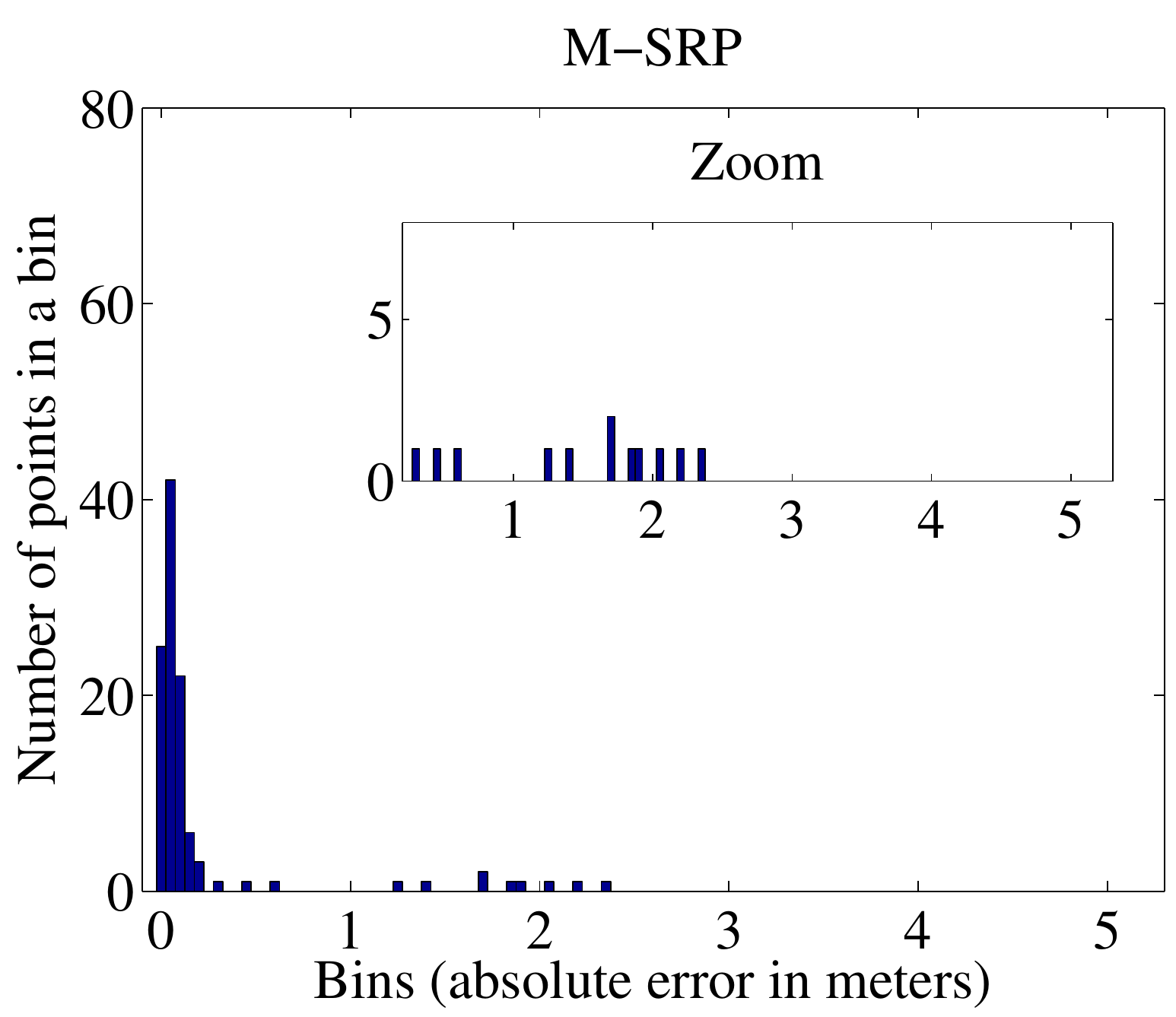}
\label{fig:cobos_simu}}
\caption{Histograms of location estimation errors for the C-SRP, RV-SRP, and M-SRP (bin width is $5$~cm). {The inside histograms (Zoom) show the number of estimation errors larger than 30 cm.}\label{fig:vsrp_simu}}
\end{figure*}

Fig.~\ref{fig:vsrp_simu} shows the histogram of the estimation error for the three localization methods employed. Table~\ref{tab:artificial_summary} summarizes the results for this scenario, including the mean {and median} estimation {errors} for each method and
its approximate number of {arithmetic} operations.
The C-SRP with a 1-cm resolution was included in order to illustrate its demanding computational complexity, even though this resolution was not used in the evaluation.
\begin{table*}[!t]
\caption{Results for the {simulated} scenario.}\label{tab:artificial_summary}
\centering
{
\begin{tabular}{|c|C{1.8cm}|C{1.8cm}|C{1.8cm}|C{1.8cm}|c|}
\hline
\multirow{2}{*}{Method [grid resolution]} & \multicolumn{2}{ c| }{Error [cm] for T60 $=$ 250~ms} & \multicolumn{2}{ c| }{Error [cm] for T60 $=$ 500~ms}  & \multirow{2}{*}{Approx. number of op. per frame ($\times 10^7$)}\\
\cline{2-5}
       & Mean & Median & Mean & Median &  \\
\hline
C-SRP  [$1$~cm]                 &    --     &  -- &    --     &  --  & $863$  \\
C-SRP  [$3$~cm]                 &  $4.87$  & $3.56$ &  $63.15$  & $6.23$ & $32.4$  \\
RV-SRP [$10$~cm, $64$~pt / ref. $1$~cm]     &  $5.04$  & $2.33$ &  $9.76$  & $2.86$ & $4.59$
  \\
M-SRP  [$10$~cm]                &  $7.91$  & $7.85$ &  $24.30$  & $9.41$ & $4.98$  \\
\hline
\end{tabular}
}
\end{table*}

%

{By observing Figs.~\ref{fig:3cmsrp_simu_250},~\ref{fig:10cmvsrp_simu_250}, and~\ref{fig:cobos_simu_250} one can verify
that all algorithms are able to localize the acoustic source at the positions tested with a relatively high accuracy when the reverberation time is moderately low (T60~$=$~250~ms). Indeed, the zoom plots show that those methods do not yield any anomalous estimate,
which is reflected in the close mean and median estimation error values in the respective column of Table~\ref{tab:artificial_summary}. Moreover, when  computational complexity of the methods is also taken into account, one can observe from these results that the proposed RV-SRP algorithm achieves the best trade-off between performance and computational complexity in this particular environment.}

{Regarding the environment with T60~$=$~500~ms, in} the case of the C-SRP method {with grid resolution of $3$~cm},  its mean estimation error was approximately {$63.15$~cm}, even though the majority of
the absolute errors were smaller than $15$~cm (see Fig.~\ref{fig:3cmsrp_simu}). If the designer of the system is somehow able to discard most of the anomalous estimates, then  the average value of the estimation error would be obviously smaller. For instance, the median estimation error of this example is around {$6.23$}~cm.
This result is achieved by performing about {$32.4\times 10^7$} {arithmetic} operations per frame.
Even after this significant reduction as compared to the $1$-cm resolution case, the computational complexity might be still too high for the envisaged application.

When using the RV-SRP, the estimation error value
is drastically reduced to approximately {$9.76$}~cm, while its median value is around {$2.86$}~cm.
{Note that the zoom plot of the RV-SRP in Fig.~\ref{fig:10cmvsrp_simu} shows that the proposed method yields very few anomalous estimates in this particular setup.}
In addition to this significant
performance enhancement, the total number of {arithmetic} operations per frame required by the functional evaluations of the proposed method is
around {$4.59\times10^7$},
about one order of magnitude smaller than the C-SRP algorithm with $3$-cm grid resolution.
It is worth pointing out that such improvements obtained by the proposed algorithm
come at the price of spending more memory resources.

The M-SRP obtained mean and median estimation errors of approximately $24.30$~cm and $9.41$~cm, respectively, both higher than those obtained by the RV-SRP method.
Besides, the number of {arithmetic} operations required by the functional evaluations of the M-SRP is around
$4.98\times10^7$ {per frame, higher than the computational cost of the method
proposed in this paper, for this particular scenario.}

\subsection{More on the Relation Between Point and Volumetric Grids}\label{sub:rel_grids}


{In the results shown in the previous subsections, volumes with $10$~cm of edge were used.
Inside each volume there were $16$ points for the real-data (2-D) scenario and
$64$ points for the simulated (3-D) scenario, thus implying that in both cases the number of
points per edge was $4$ (refer to Figs.~\ref{fig:grid} and~\ref{fig:vols}).
For these two scenarios, an increase in the number of points per edge, and thus per volume,
did not lead to a significant performance gain that would justify an increase in the computational
complexity.
Hence, $4$ points per edge proved to be a good choice when using volumes with $10$~cm of edge.
}

{If one intends to use the proposed methods with coarser volumetric grids, i.e. using volumes
with larger edges, then one should be aware that the complexity of the RV-SRP is not a monotonic
decreasing function of the size of the volume.
For instance, the number of {arithmetic} operations required by the RV-SRP with $100$~cm of grid resolution (edge) is
higher than the one with $10$~cm; besides the results are much worse due to the coarser
volumetric grid.
Experimental observations have pointed out that even when using volumes
with large edges, such as $50$ and $100$~cm, $8$ points per edge were enough. This choice depends
on several parameters, among them the sampling frequency (in this section, 48~kHz).

\section{Concluding Remarks}\label{sec:conclusion}

This paper introduced a novel approach to the application of the SRP method to the problem of sound source
localization using microphone arrays. In order to tackle high resolution requirements without resorting
to a superfine grid, which would lead to an exceedingly complex procedure, the proposed V-SRP performs
the search over a sparse volumetric grid; the volume with the highest {objective function} value is expected to
contain the sound source. Its variant, the RV-SRP, further refines the search by applying the classical SRP method
inside the winning volume. Complexity analysis as well as two sets of experiments (2-dimensional search
using uniform linear array and natural signals, and 3-dimensional search using planar array and simulated
signals) demonstrate that the V-SRP and the RV-SRP outperform the classical SRP and another recent competing
method (the M-SRP), achieving a comparable accuracy with much reduced complexity.

The proposed approach provides some degrees of freedom that can be customized for a given application.
For instance, one can use other volumes different from cubes, with variable sizes---possibly chosen
with the aid of a discriminability measure~\cite{Lonnes_iwaenc2012}---or perform the refinement step
using a source localization method other than the classical SRP. In addition, {when} accuracy is
of paramount importance and computational resources are abundant (e.g. in cloud computing cases), the
RV-SRP method can be adjusted to retain not just a single winning volume, but the $N$ volumes that lead
to the highest {objective function} values; their respective refinement steps can then be performed in a
distributed fashion.

\section*{Acknowledgment}
This R\&D project resilted from a cooperation between Hewlett-Packard Brasil Ltda. and COPPE/UFRJ, being supported with resources of Informatics Law (no. 8.248, from 1991).
L. W. P. Biscainho, T. N. Ferreira, M. V. S. Lima, W. A. Martins, and L. O. Nunes would like to thank also CAPES, CNPq, and FAPERJ agencies for funding their research work.

\section*{Note}
An almost identical version of this manuscript was submitted to the IEEE Transactions on Audio, Speech, 
and Language Processing (TASLP) in June 2013, and eventually rejected in May 2014. 
Since one of the reviewers' concerns was the paper length, the authors decided to reshape the work as a letter 
to be submitted to the IEEE Signal Processing Letters, while at the same time making the longer version, 
which discusses e.g. computational requirements, available.



%
%
%
%

\bibliographystyle{IEEEtran}
\bibliography{bib-v7}

%
%
%
%
%
%
%

\clearpage

\begin{center}
 {\Large SUPPLEMENTARY MATERIAL}
\end{center}

\begin{abstract}
 The real-data and simulated scenarios presented in Section~\ref{sec:results} are revisited and results using 
 other competing algorithms are included. 
 These new algorithms are the stochastic region contraction (SRC)~\cite{Silverman2007} and 
 the Iterative SRP-based method (I-SRP)~\cite{Marti2013}. 
 
 It is important to highlight that this part of the text contains new results, which were not part of the 
 original manuscript submitted to the IEEE TALSP, that supplement the paper we submitted to IEEE Signal Processing Letters 
 entitled ``A Volumetric SRP with Refinement Step for Sound Source Localization''.
\end{abstract}

\section{Additional Results}

\begin{table*}[H]
\caption{{Results for the real-data scenario.}}\label{tab:NEW_real_data_summary}
\vspace{-0.3cm}
\centering
{
\begin{tabular}{|c|c|c|c|}
\hline
Method [grid resolution]        & Mean error [cm] & Median error [cm] & Approx. number of op. per frame ($\times 10^5$) \\
\hline
C-SRP  [$1$~cm]                 &  $19.62$  &  $3.15$  &  $38.0$  \\
C-SRP  [$10$~cm]                &  $52.09$  &  $11.17$  &  $0.398$  \\
M-SRP  [$10$~cm]                &  $19.67$  &  $6.76$  &  $2.71$  \\
I-SRP  [$10$~cm ($1$~iteration)]            &  $38.69$  &  $7.87$   &  $3.54$     \\
I-SRP  [$10$~cm, $1$~cm ($2$~iterations)]   &  $36.24$  &  $6.05$   &  $3.65$     \\
I-SRP  [$50$~cm, $10$~cm, $1$~cm ($3$~iterations)]   &  $130.78$  &  $85.74$   &  $0.77$     \\
SRC    [$10$~cm]                            &  $47.99$  &  $8.99$   &  $7.00$     \\
\hline
V-SRP  [$10$~cm, $16$~pt]                   &  $18.29$  &  $6.21$  &  $2.08$  \\
RV-SRP [$10$~cm, $16$~pt / ref. $1$~cm]     &  $15.98$  &  $3.77$  &  {$2.11$} \\
\hline
\end{tabular}
}
\end{table*}

\begin{table*}[!b]
\caption{{Results for the simulated scenario.}}\label{tab:NEW_artificial_summary}
\vspace{-0.3cm}
\centering
{
\hspace{-15.7mm}
\begin{tabular}{|c|C{1.8cm}|C{1.8cm}|C{1.8cm}|C{1.8cm}|c|}
\hline
\multirow{2}{*}{Method [grid resolution]} & \multicolumn{2}{ c| }{Error [cm] for T60 $=$ 250~ms} & \multicolumn{2}{ c| }{Error [cm] for T60 $=$ 500~ms}  & \multirow{2}{*}{Approx. number of op. per frame ($\times 10^7$)}\\
\cline{2-5}
       & Mean & Median & Mean & Median &  \\
\hline
C-SRP  [$1$~cm]                                     &    --     &     --   &    --     &  --  & $863$  \\
C-SRP  [$3$~cm]                                     &  $4.87$   &   $3.56$ &  $63.15$  & $6.23$  & $32.4$  \\
M-SRP  [$10$~cm]                                    &  $7.91$   &   $7.85$ &  $24.30$  & $9.41$  & $4.98$  \\
I-SRP  [$10$~cm ($1$~iteration)]                    &  $68.96$  &  $14.32$ & $166.40$  & $18.94$ & $6.84$  \\
I-SRP  [$10$~cm, $1$~cm ($2$~iterations)]           &  $65.01$  &   $9.41$ & $164.16$  & $12.91$ & $6.98$  \\
I-SRP  [$50$~cm, $10$~cm, $1$~cm ($3$~iterations)]  & $285.07$  & $320.13$ & $337.99$  &$374.26$ &   --    \\
SRC    [$10$~cm]                                    &  $86.42$  &   $3.98$ & $192.89$  &$198.88$ & $0.5$\\
\hline
V-SRP [$10$~cm, $64$~pt]                            &   $9.88$  &   $7.55$ &  $14.41$  &  $9.95$ & $< 4.59$  \\
RV-SRP [$10$~cm, $64$~pt / ref. $1$~cm]             &   $5.04$  &   $2.33$ &   $9.76$  &  $2.86$ & $4.59$  \\
\hline
\end{tabular}
}
\end{table*}

In this section, we summarize the additional results for the real-data scenario (refer to Section~\ref{sub:exper}) 
and for the simulated scenario (refer to Section~\ref{sub:artif}). 
The new competing algorithms are the SRC~\cite{Silverman2007} and the I-SRP~\cite{Marti2013}.
The SRC evaluated $3000$ points per volume among which $100$ were chosen to define the new volume 
employed on the next iteration; 
this process was repeated until the volume's edge achieved $10$~cm.
As for the I-SRP, three configurations were  used: 
\begin{enumerate}
 \item I-SRP with a single iteration and grid resolution of $10$~cm: this allows a fair comparison between the M-SRP and the I-SRP, 
 which were proposed by the same group;
 \item I-SRP with two iterations (first iteration with grid resolution of $10$~cm and second iteration with 
 grid resolution of $1$~cm): this allows a fair comparison between the RV-SRP and the I-SRP;
 \item I-SRP with three iterations (first iteration with grid resolution of $50$~cm and last iteration with 
 grid resolution of $1$~cm): this is the configuration used in~\cite{Marti2013}. 
\end{enumerate}

The results for the real-data and simulated scenarios are summarized in Tables~\ref{tab:NEW_real_data_summary} 
and~\ref{tab:NEW_artificial_summary}, respectively.

In summary, both the SRC and the I-SRP yielded inferior results, especially when facing large rooms and/or 
higher reverberation. 
In these experiments, the proposed V-SRP and RV-SRP were the most robust methods with respect to the different 
array geometries, reverberation time, and room dimension.

\end{document}